\documentclass{aa}  
\usepackage{graphicx}
\usepackage{txfonts}
\usepackage{float}
\begin{document}

   \title{Spatially resolved spectroscopy across stellar surfaces. II. }

   \subtitle{High-resolution spectra across HD~209458 (G0 V) }

   \author{Dainis Dravins
          \inst{1},
               Hans-G\"{u}nter Ludwig
           \inst{2},
          Erik Dahl\'{e}n,
        \inst{1}
          \and
        Hiva Pazira
          \inst{1,3} \fnmsep 
                }
%
%
\institute{Lund Observatory, Box 43, SE-22100 Lund, Sweden\\
              \email{dainis@astro.lu.se}
\and
     Zentrum f\"{u}r Astronomie der Universit\"{a}t Heidelberg, Landessternwarte K\"{o}nigstuhl, DE--69117 Heidelberg, Germany\\
              \email{hludwig@lsw.uni-heidelberg.de}
\and 
                Present address: Department of Astronomy, AlbaNova University Center, SE--10691 Stockholm, Sweden\\
             }

   \titlerunning{Spatially resolved stellar spectroscopy. II.}
   \authorrunning{D.\ Dravins et al.} 
 
\date{Received XXX YY, 2017; accepted XXX YY, 2017}

 
\abstract
   {High-resolution spectroscopy across spatially resolved stellar surfaces aims at obtaining spectral-line profiles that are free from rotational broadening; the gradual changes of these profiles from disk center toward the stellar limb reveal properties of atmospheric fine structure, which are possible to model with 3-D hydrodynamics.}
   {Previous such studies have only been carried out for the Sun but are now extended to other stars.  In this work, profiles of photospheric spectral lines are retrieved across the disk of the planet-hosting star HD~209458 (G0~V).}
   {During exoplanet transit, stellar surface portions successively become hidden and differential spectroscopy provides spectra of small surface segments temporarily hidden behind the planet.  The method was elaborated in Paper I, with observable signatures quantitatively predicted from hydrodynamic simulations. }
   {From observations of HD~209458 with spectral resolution $\lambda/\Delta \lambda\sim$80,000,  photospheric Fe~I line profiles are obtained at several center-to-limb positions, reaching adequately high S/N after averaging over numerous similar lines.}
   {Retrieved line profiles are compared to synthetic line profiles.  Hydrodynamic 3\mbox{-}D models predict, and current observations confirm, that photospheric absorption lines become broader and shallower toward the stellar limb, reflecting that horizontal velocities in stellar granulation are greater than vertical velocities.  Additional types of 3\mbox{-}D signatures will become observable with the highest resolution spectrometers at large telescopes. }


\maketitle

\section{Spatially resolved stellar spectra}

Three-dimensional and time-dependent hydrodynamic simulations provide realistic descriptions of the atmospheres of various classes of stars, and spectra computed from such models can be used to determine precise properties of the star and its exoplanets.  To constrain and evolve such models, observations beyond the ordinary spectrum of integrated starlight are desirable.  Spectral-line syntheses in 3\mbox{-}D atmospheres show a rich variety of phenomena characterizing stellar hydrodynamics, which are seen especially in the gradual changes of photospheric line profile strengths, shapes, asymmetries, and wavelength shifts from the center of the disk toward the stellar limb.  However, direct comparisons between theory and spectral-line observations have in the past only been possible for the spatially resolved Sun  \citep[e.g.,][]{lindetal17}. In Paper~I \citep{dravinsetal17}, we examined theoretically predicted spatially resolved signatures for a group of main-sequence stellar models with temperatures between 6730 and 3960 K.  Corresponding observations are feasible during exoplanet transits when small stellar surface portions successively become hidden, and differential spectroscopy between different transit phases can provide spectra of small surface segments temporarily hidden behind the planet.  The observational requirements were elaborated in Paper~I, in which observable parameters were predicted quantitatively. 

In this Paper~II, an observational demonstration of this method is presented.  We first examine the existence of suitable exoplanet transit stars and the availability of observational data for those stars.  As a first target, the G0~V star HD~209458 is selected, whose T$_{\textrm{eff}}$ $\sim$6000 K is near the middle of the model interval treated in Paper~I.  Spectra of exceptionally high signal to noise are required; methods for obtaining such spectra, and for verifying the data integrity for the extraction of spatially resolved data, are explained.  Reconstructed line profiles at different positions across the stellar disk are presented and compared to synthetic spectra from hydrodynamic models with parameters close to those of HD~209458.  Finally, the potential for more exhaustive future observations is outlined.

\section{Candidate stars with transiting planets} 

\begin{figure}[H]
\centering
\includegraphics[width=\hsize]{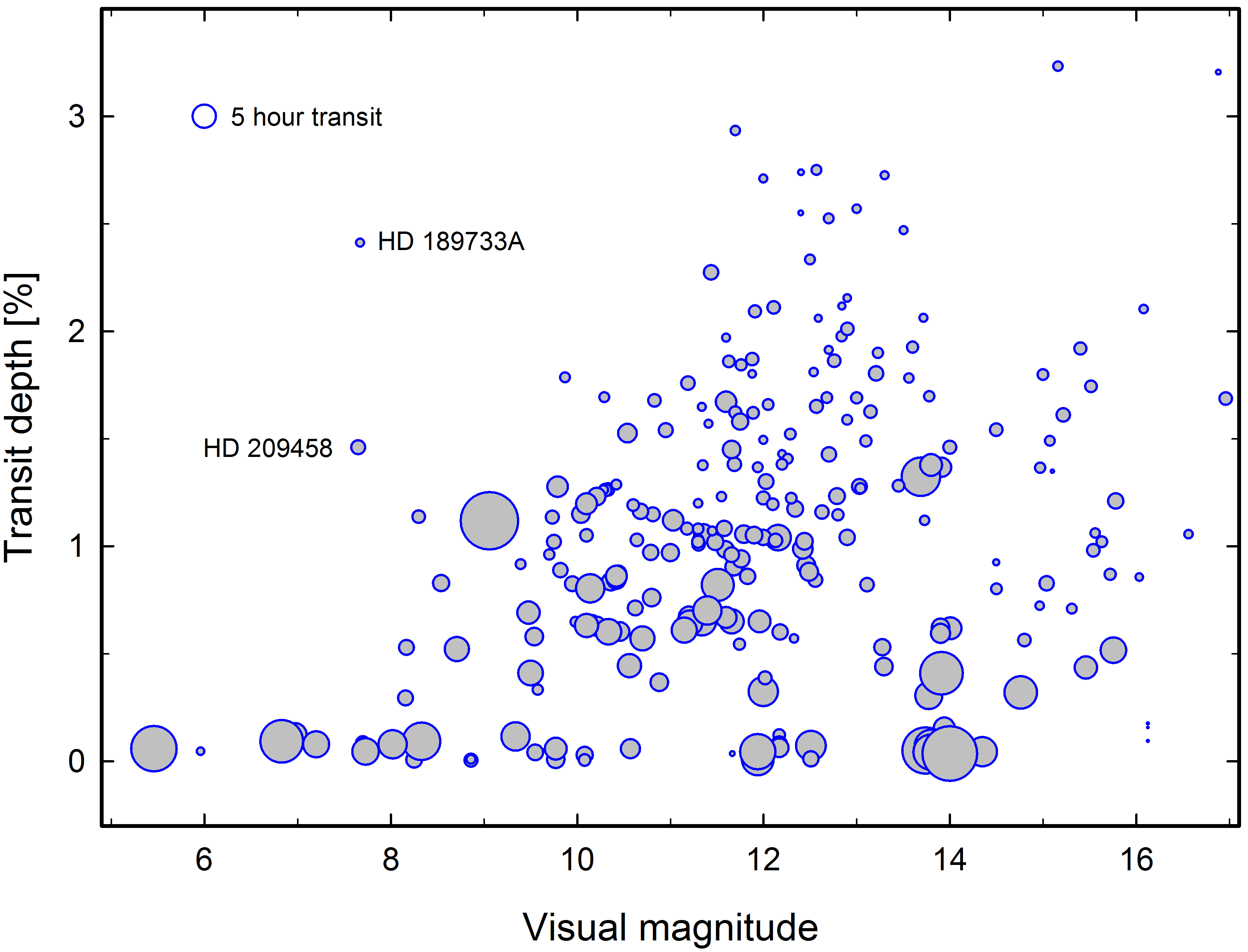}
\caption{Photometric transit depth vs. apparent stellar m$_{\textrm{V}}$ magnitude for transiting exoplanet systems.  Symbol diameters are proportional to the duration of transit.  The demanding observational requirements limit current studies to the brightest stars with the largest planets, i.e., HD~209458 and HD~189733A.  Data for systems with reasonably well-determined properties were taken from the \citet{exoplanets17}; \citet{hanetal14}. }
\label{fig:planet_statistics}
\end{figure}

As emphasized in Paper~I, the method is observationally challenging since exoplanets cover only a tiny fraction of the stellar disk, and only spectra with exceptionally low noise permit the reconstruction of any sensible spatially resolved data.  Figure~\ref{fig:planet_statistics} shows the distribution of currently known stellar systems with transiting planets for different apparent brightness, planet size, and transit duration.  Realistically, only Jupiter-size (or larger) planets transiting main-sequence dwarf stars can be targeted for study (subtending at least $\sim$1\,\% of the stellar disk).  The requirements limit these studies to the brightest host stars with the largest planets, in particular HD~209458 (G0~V) and HD~189733A (K1~V), and these were selected for detailed investigations.  The first of these is the topic of this paper.

\subsection{HD~209458 and its exoplanet Osiris}

HD~209458 was the first star with an observed planetary transit \citep{charbonneauetal00, henryetal00}; this star has since been extensively studied from the ground and from space.  Its bloated `hot Jupiter' with R$_{\textrm{p}}$ $\sim$1.4 R$_{\textrm{Jup}}$ is about as large as planets can get, offering a photometric transit depth of $\sim$1.5\,\% at visual wavelengths \citep{brownetal01}.  While the transit path does not reach the very center of the stellar disk, it extends to almost $\mu$ = cos\,$\theta$ = 0.9; at this position, the spectrum is not much different from the disk-center value at $\mu$ = 1. 

HD~209458 is close to the Sun in spectral type, usually classified as G0~V (sometimes F9~V).  The apparent brightness of m$_{\textrm{V}}$~=~7.65, has enabled detailed studies of both the host star and its exoplanet.  A critical examination of the stellar properties by \citet{delburgoallende16} arrived at best-fit values for the stellar radius R$_{\star}$ = 1.20 $\pm$\,0.05 R$_{\odot}$, T$_{\textrm{eff}}$ = 6071 $\pm$\,20 K, log~$\varg$ [cgs] = 4.38, [Fe/H] = 0, and [$\alpha$/H] = 0.  The spectrum of HD~209458 is sufficiently similar to the solar spectrum to enable rather straightforward line identifications using solar spectrum line lists, thanks to both its closeness in temperature and because it also is a slow rotator ($V_\textrm{rot}$ $\sim$\,4 km\,s$^{-1}$; sin\,$i$ = 1 if the star rotates in the plane of its transiting planet).  However, because HD~209458 is a little hotter, its photospheric lines are typically somewhat weaker than solar photospheric lines which, however, also means less blending. 

The exoplanet has been called `Osiris' and, for simplicity, we also use this name.  An evaluation of its radius gives R$_{\textrm{p}}$ = 1.41 $\pm$\,0.06 R$_{\textrm{Jup}}$ \citep{delburgoallende16}.  Numerous studies of its atmosphere, including its extended hydrogen and sodium exosphere and of atmospheric dynamics have been made; for summaries see \citet{demingseager17}, \citet{fossatietal15}, \citet{lineetal16}, and \citet{perryman11}.  Daytime temperatures of $\sim$1,400--1,800 K \citep[e.g.,][]{lineetal16}, imply an extended atmosphere with layers of light gases and molecular constituents such as H$_{\textrm{2}}$O, CH$_{\textrm{4}}$, and CO$_{\textrm{2}}$.  While this enables the study of the atmospheric chemistry and mass loss of the planet, the presence of spectral lines from such planetary species and the different effective planet sizes in the corresponding monochromatic light imply that reconstructions of stellar spectra at the precise wavelengths corresponding to such molecular lines could be a more complex task. We do not  attempt this task here.  Rather, this study is about stellar photospheric Fe~I lines, whose formation requires temperatures that are much higher than the planetary temperatures, and where the optical density of the upper planetary atmosphere (and thus the effective planet size) does not change across such lines.  At such wavelengths, it must be a closely valid approximation to treat the planet as an opaque body, whose size is indicated by transit photometry in adjacent passbands \citep{deegetal01, jhaetal00}.

\subsection{Osiris orbit and transit geometry}

The geometry of the exoplanet transit must be known to fully interpret any observed line-profile variations.  Through the Rossiter-McLaughlin effect, the position and path of the planet across the stellar disk can be determined.  Details of that method were explained by \citet{gaudiwinn07}, \citet{gimenez06}, and \citet{hiranoetal10, hiranoetal11}, and applied specifically to HD~209458 and Osiris by \citet{ohtaetal05}, \citet{quelozetal00}, \citet{snellen04}, \citet{torresetal08}, \citet{winnetal05}, and \citet{wittenmyeretal05}.  

Different authors arrive at slightly different values for the various parameters, but agree that Osiris moves in an almost circular prograde orbit (eccentricity <\,0.02), of period 3.52 days, with an inclination around $87^{\circ}$.  The orbital radius of 0.047 au implies an impact parameter $\sim$0.51 (minimum sky-projected distance to stellar disk center in units of R$_{\star}$).  The maximum photometric transit depth reached during its 3.1-hour transit is 1.46$\,\%$ in the visible \citep{exoplanets17, extrasolarplanets17}.

\section{Observational requirements}

The exceptional signal-to-noise requirements limit usable data to the highest fidelity spectra from high-resolution spectrometers at very large telescopes.  Fortunately, any realistic, bright target star is among the very objects of which extensive observations have already been carried out to characterize the exoplanet and its atmosphere; numerous spectra of these objects are available in observatory archives.

\subsection{Observations of HD~209458}

For HD~209458, hundreds of archive spectra were retrieved from several different observatories and examined for their suitability.  In particular, the ESO Science Archive Facility ({\url{http://archive.eso.org/}}) was examined for data from the UVES and HARPS spectrometers at the VLT Kueyen on Paranal and the 3.6 m telescope on La Silla, respectively; the Keck Observatory Archive ({\url{http://www2.keck.hawaii.edu/koa/}) for data from its High Resolution Echelle Spectrometer (HIRES) at Keck-1 on Maunakea; and the SMOKA science archive (Subaru Mitaka Okayama Kiso Archive; {\url{http://smoka.nao.ac.jp/}}) for spectra from the High Dispersion Spectrograph (HDS) at the Subaru telescope, also on Maunakea. 

A sequence of test analyses of data from different instruments revealed the limitations of most, even otherwise excellent, high-resolution spectra.  However, a few observing campaigns had been carried out with somewhat lessened spectral resolution but aiming at the highest possible photometric precision.  When the S/N in individual spectral recordings reaches $\sim$\,400 or more, a threshold seems to be reached, where data start to become useful, even though extensive averaging of many spectral lines and of multiple exposures is still required.

\begin{figure}[H]
\centering
\includegraphics[width=\hsize]{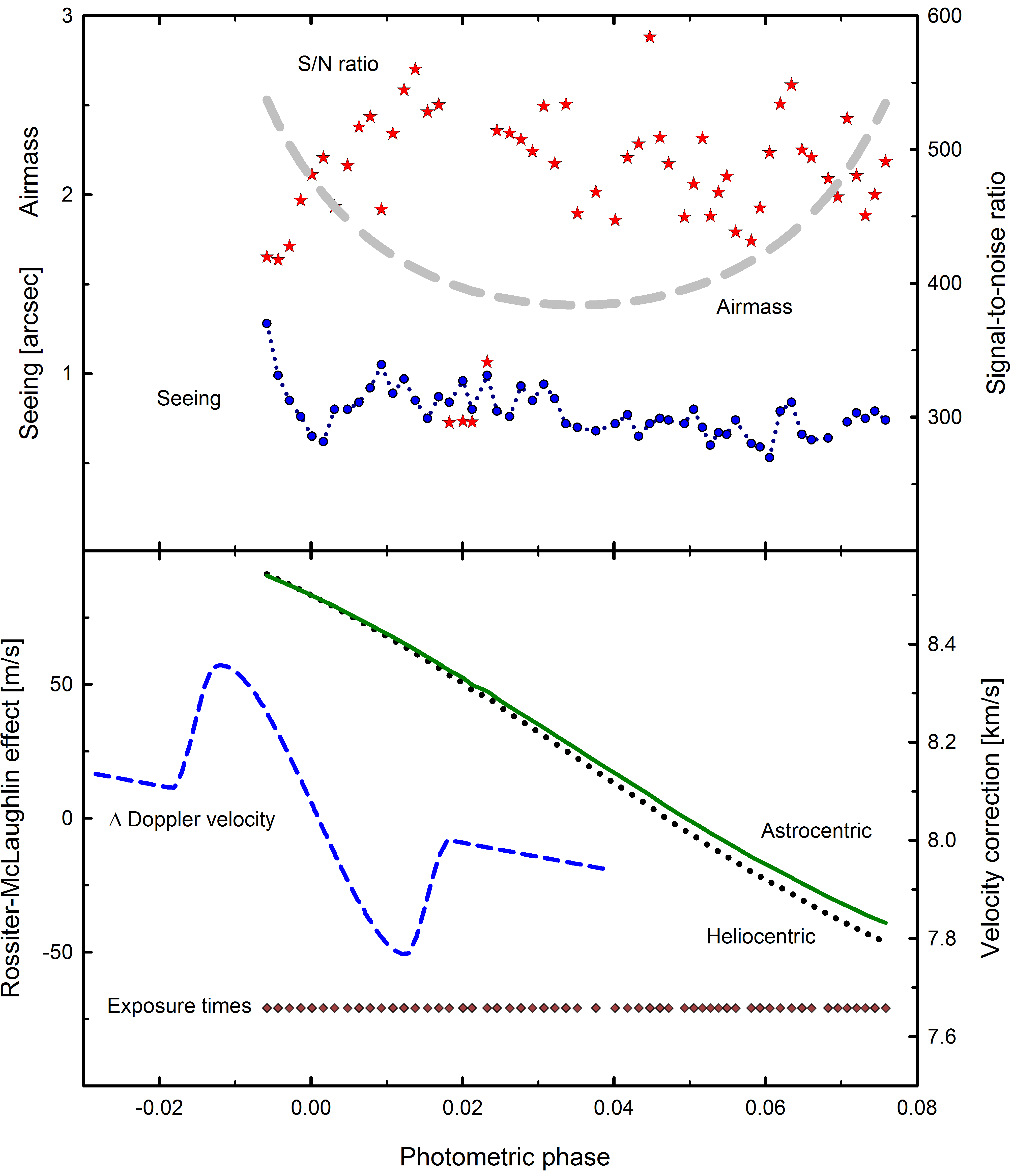}
\caption{Parameters for the UVES observations.  The photometric S/N refers to the best-exposed echelle order as computed by the data reduction pipeline.   Passing clouds caused a drop around orbital phase 0.02 (photometric phase is zero at transit mid-point during the 3.52-day orbit of the planet).  Data from the Paranal seeing monitor show image stability despite substantial airmass changes during the night.  The exposure starting times are for the REDL part of the UVES spectrum; the sequence started with the transit already in progress.  The heliocentric velocity correction concerns the velocity vector between the observatory and the direction toward the star, while the astrocentric correction also takes into account the barycentric motion of HD~209458 induced by its planet.  The Rossiter-McLaughlin signature during the transit was adapted from \citet{winnetal05} and \citet{wittenmyeretal05}. }
\label{fig:metadata}
\end{figure}

The data selected for the current analysis originate from one observing night with the UVES spectrometer \citep{dekkeretal00} on August 14, 2006 at the ESO VLT Kueyen telescope on Paranal, originally recorded for an investigation of the atmosphere of Osiris, for which results were documented in \citet{albrechtetal09} and \citet{snellenetal11}. The three UVES detectors yield spectra in a shorter wavelength BLUE region, in a longer wavelength visual REDL, and in REDU toward the near-infrared.  The signal-to-noise values for these 400-second exposures, as computed by the ESO UVES data analysis pipeline, often exceed 500 in the centers of the best-exposed echelle orders of the REDL part of the spectrum.  Such low noise levels in these deep exposures were enabled by widening the spectrometer entrance slit to 0.5 arcsec at the cost of then reducing the spectral resolution to $\lambda$/$\Delta\lambda$ $\approx$~80,000 in the red arm of the UVES spectrometer versus\ $\sim$110,000 in its more common setup with an 0.3 arcsec slit.  Figure~\ref{fig:metadata} summarizes conditions during the observing night while Fig.~\ref{fig:transit_geometry} shows the transit geometry.  The limb-darkening functions shown are those computed specifically for HD~209458 by \citet{hayeketal12} in the two passbands SDSS $g'$ (blue) and SDSS $r'$ (red).  Their effective wavelengths, 475.1 and 620.4 nm \citep{fukugitaetal96}, closely match our spectral-line selections.

\begin{figure}[H]
\centering
\includegraphics[width=85 mm]{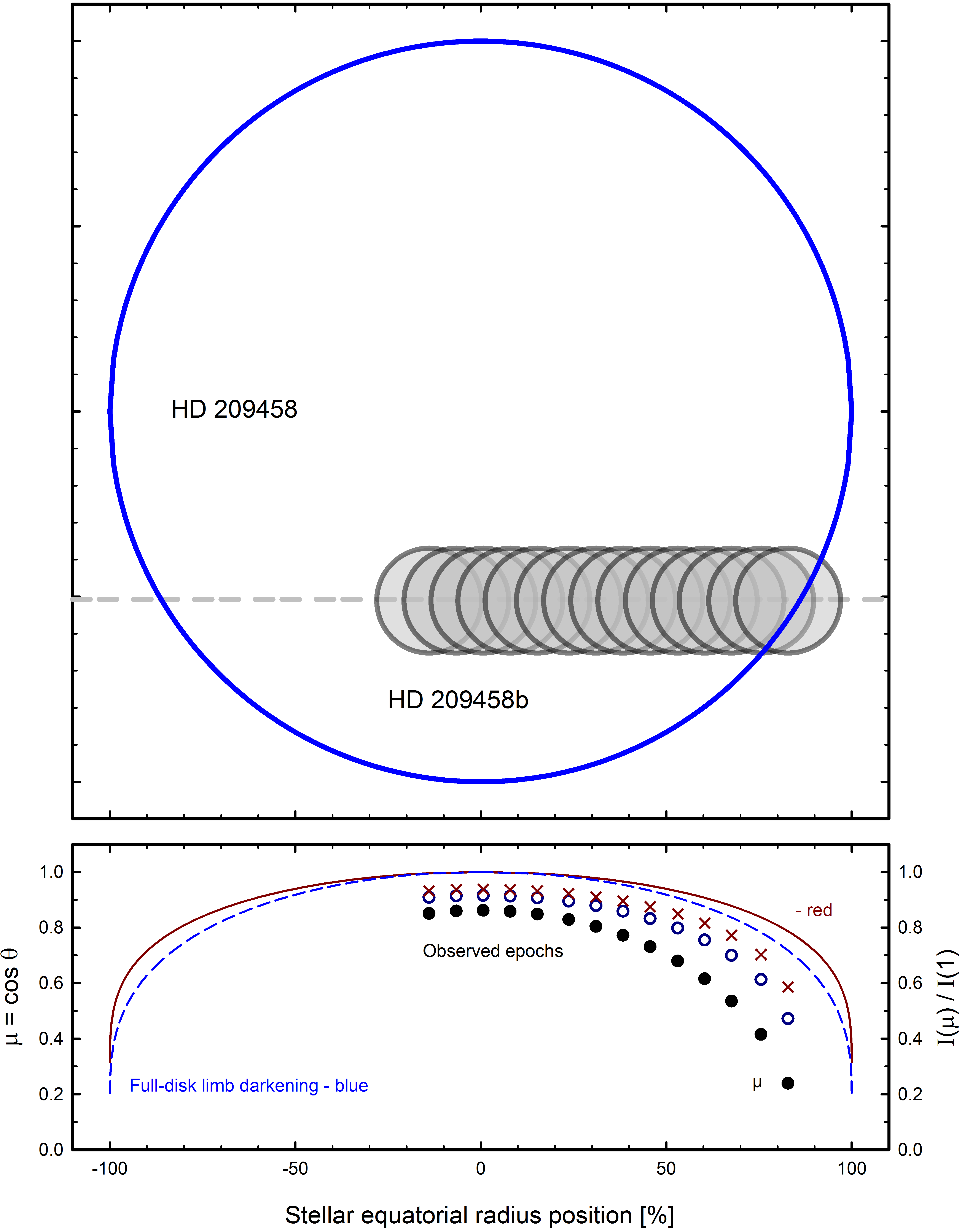}
\caption{Exoplanet transit geometry during the observing night.  The planet size and positions during successive REDL spectral exposures are to scale.  The bottom panel shows curves for the theoretical limb-darkening functions in red and blue for the full disk, along the stellar equator (scales at right and bottom).  Solid dots show the stellar center angle $\mu$ = cos\,$\theta$ at the observed epochs (scales at left and bottom, also top panel), with $\mu$ ranging between 0.86 and 0.24.  These values lie below the full-disk curves since the transit does not reach the disk center.  For these observed epochs, the limb-darkening values in blue (open circles) and red (crosses) are shown as ratios between intensity at the actual planet location and disk center (scales at right and bottom).  At the final epochs, part of the planet is already outside the stellar disk.}
\label{fig:transit_geometry}
\end{figure}

\subsection{Exoplanet transit ephemeris}

Orbital parameters and transit epochs are compiled in databases such as the \citet{exoplanets17, extrasolarplanets17}, or \citet{nasa17}.  The present event had mid-transit at 2453961.59757927 heliocentric Julian Date.  A full transit duration is $\sim$188 min and, at the stellar apparent latitude of $27^{\circ}$, the length of the transit path equals 0.89 stellar diameters; the transit proceeds with 0.0047 stellar diameters/min or 1 planet diameter per 25 minutes.  With the planet diameter =~0.12\,$\times$\,stellar, 0.26 planetary diameters are traversed during each 400 s exposure and, around mid-exposure, starlight is gradually obscured in an interval between $\pm$ 0.13 planet diameters, which is equal to $\pm$1.6\,\% of the stellar diameter.  Other than very close to the limb, the effects of such smearing are largely negligible (Fig.\ \ref{fig:transit_geometry}).

\subsection{Barycentric motion of the host star}
 
The orbital eccentricity of the planet is very low, i.e., not significantly different from zero.  Osiris induces a smooth variation in the barycentric motion of the star, whose radial velocity can be closely approximated by a sinusoidal curve with amplitude 85 m\,s$^{-1}$ \citep{laughlinetal05, winnetal05, wittenmyeretal05}. The deviation from such a sine curve (except for the Rossiter-McLaughlin event during the transit itself) is no more than some m\,s$^{-1}$, well inside the error bars.  Such a function was applied for the correction of heliocentric wavelength scales to astrocentric values (Fig.~\ref{fig:metadata}).

\subsection{Photometric transit}

For the later spectral reconstruction, photometric data are required to know the amount of flux temporarily obscured by the planet.  Since the transits are repetitive, such data may be from a different epoch than the spectroscopic measurements.  \citet{brownetal01} obtained precise transit photometry with the STIS spectrograph on the Hubble Space Telescope.  Their time series of 80~s sampling intervals achieved a relative precision of $\sim$10$^{-4}$ per sample.  The wavelength interval of 582--638 nm closely overlaps with our UVES REDL data, and their measurements are used in the calculations below.

\begin{figure}[H]
\centering
\includegraphics[width=\hsize]{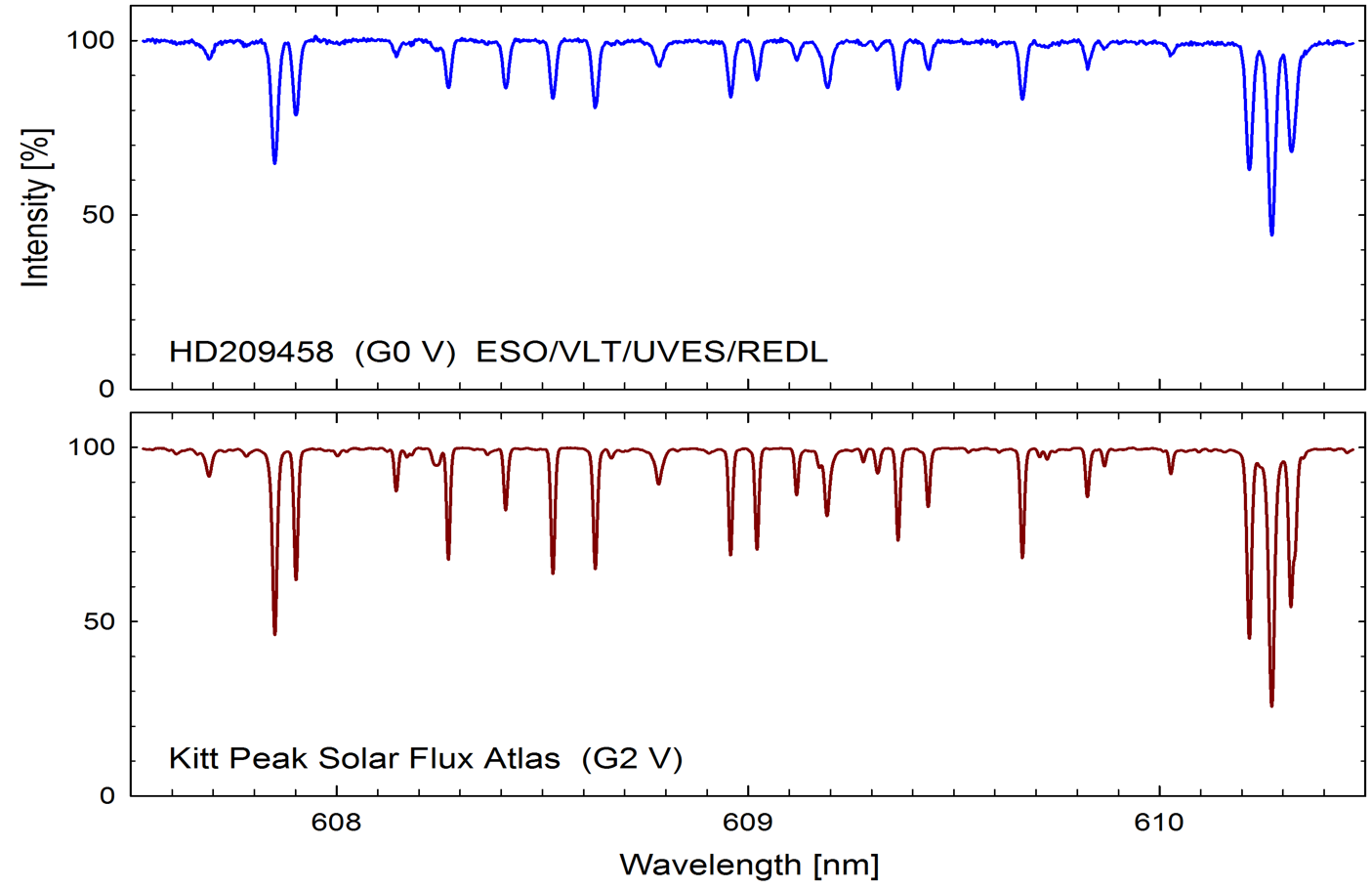}
\caption{Spectrum of the G0~V star HD~209458 is very similar to the solar spectrum \citep{kuruczetal84}, often enabling straightforward line identifications.  The stellar spectrum is here shifted in wavelength to match solar values. }
\label{fig:solar_spectrum}
\end{figure}

\section{Selecting spectroscopic data}

The full exploitation of spatially resolved stellar spectroscopy could gainfully utilize data over broad wavelength regions, of very high resolution, recorded during short intervals of the transit, with very accurate wavelength calibrations, and signal-to-noise ratios even on the order of 10,000, which are comparable to solar spectral atlases.  Since such observations are not yet feasible, we evaluated methods toward at least partially reaching that goal.  Such S/N ratios cannot be reached for individual spectral lines, and thus averaging multiple lines spread over wavelength is required.  Since usable spectral regions must embrace numerous reasonably clean lines while they are not overly disturbed by telluric features, this dictates spectral regions in the visual.  Toward the ultraviolet there is too much line blending and  there are too many telluric lines from water vapor and other species toward the infrared \citep{smetteetal15}.

\subsection{Selecting spectral lines}

With a temperature and chemical abundance close to solar, the spectrum of HD~209458 is very similar to that of the Sun, which often enables straightforward line identifications using the well-studied solar spectrum as a reference.  An example in Fig.~\ref{fig:solar_spectrum} compares one small spectral segment from current observations with that from a spectral atlas of integrated sunlight \citep{kuruczetal84}. 

To reach acceptable S/N ratios, averaging over many lines is required.  Although the spectrum contains thousands of photospheric lines, most are blended to some degree and should be avoided.  Further, to obtain cleaner signatures of atmospheric hydrodynamics, lines should preferably be from atomic species that mainly occur as one single isotope (i.e., do not have significant isotope shifts); also species that are even-even in their proton-neutron numbers (i.e., with zero nuclear spin and no hyperfine splitting) and have a large mass, minimizing thermal broadening.  The preferred species is iron, which also has a rich spectrum.  To identify lines for the later analysis, listings of apparently unblended solar lines were used.  For neutral iron, the list from \citet{stenflolindegren77} provided 410 particularly unblended Fe~I lines in the region from 400 to 686 nm, selected from the Jungfraujoch spectral atlas of the solar disk center \citep{delbouilleetal89}.  There are fewer Fe~II lines; a listing of 58 largely unblended lines was taken from \citet{dravinsetal86}.   In selecting stellar lines, the absence of close-by parasitic lines was examined in the spectrum atlases of both the solar disk center and integrated solar flux.

\begin{figure}[H]
\centering
\includegraphics[width=\hsize]{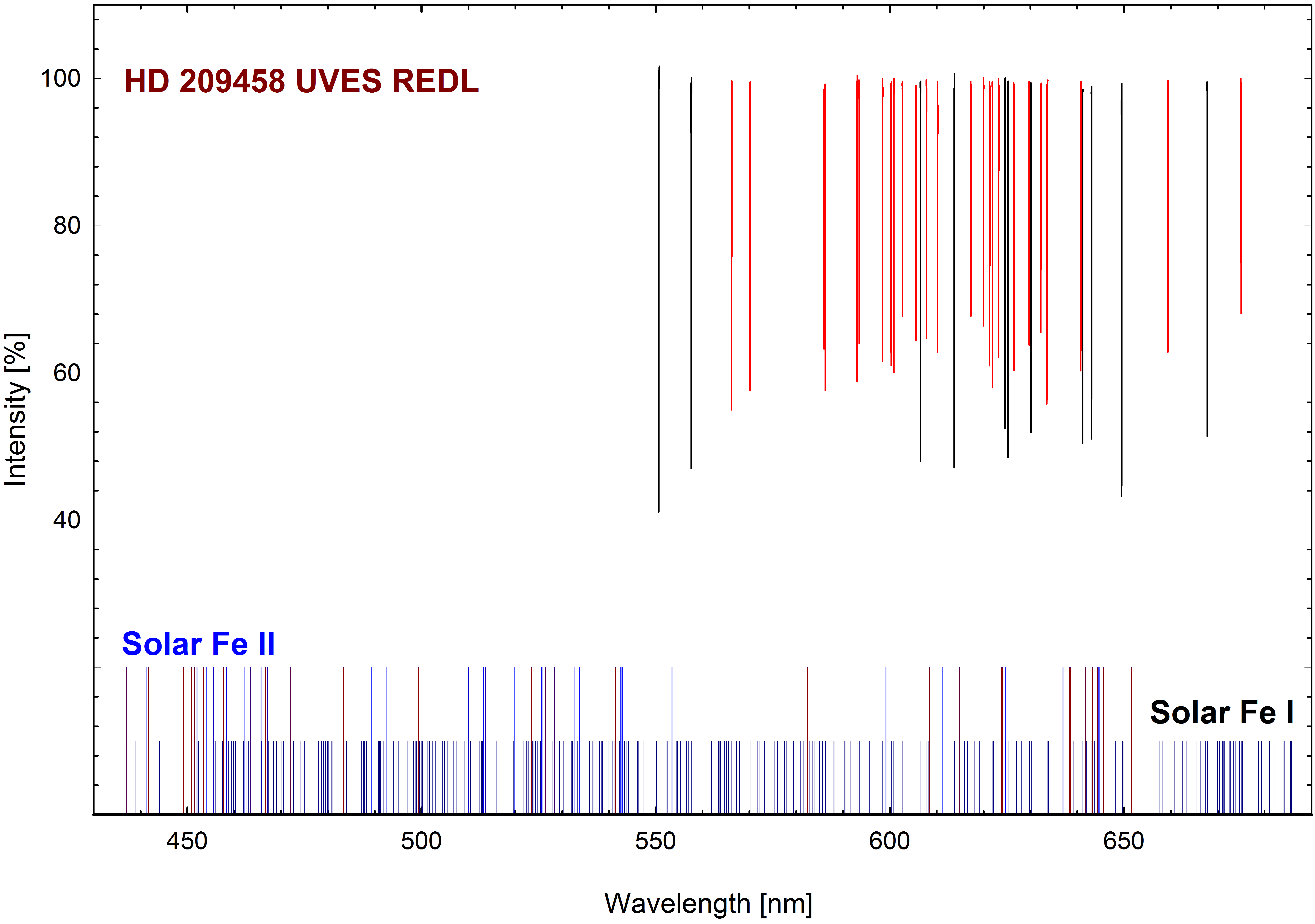}
\caption{Bottom: Wavelengths for candidate unblended Fe~I and Fe~II lines in the solar spectrum.  Top: Corresponding lines selected in HD~209458 with observed line depths marked in units of the spectral continuum.  The subgroup of finally selected 26 weaker lines is plotted in red and the 11 stronger lines are indicated in blue. }
\label{fig:line_statistics}
\end{figure}

Figure \ref{fig:line_statistics} shows wavelength positions of these candidate Fe~I and Fe~II lines.  Most of these could be identified in HD~209458 although its lines generally are weaker and broader, further smeared by the limited spectral resolution.  The UVES spectral recording covers three ranges from blue to  far red, but there are certain gaps between the ranges depending on the spectrometer settings used.  From these candidate wavelengths, some 40 apparently clean lines were initially selected in the UVES BLUE region shortward of 500 nm and some 50 lines in REDL longward of 550 nm.  Among those, a few somewhat deviant lines were removed (all from near the detector edges).  For the particular spectrometer settings used, a significant number of potential lines were lost in the gap between BLUE and REDL.  The far red region, REDU, was not searched for Fe~I lines, owing to both their paucity and the increased presence of telluric ones.  However, for the final analysis, only lines from the REDL region were ultimately retained because the lines are cleaner and the recordings are better than the photometrically noisier BLUE lines.  Also, the few Fe~II lines do not attain sufficient S/N and were not retained.  In Fig.~\ref{fig:line_statistics}, the finally retained lines in HD~209458 are indicated with both their wavelength positions and their strength plotted as observed absorption depth from the continuum.

\section{Data handling}

A total of 56 exposures had been recorded during the observing night, the sequence starting when the transit was already in progress.  Extensive analyses were made in searching for possible systematics or subtle instrumental calibration effects that might show up after massive averaging.  Specifically for UVES, such effects have also been examined by \citet{chandetal06}, \citet{molaroetal08}, \citet{snellenetal11}, \citet{thompsonetal09}, \citet{whitmoremurphy15}, and \citet{whitmoreetal10}. 

Even though UVES is a very stable instrument, it is not actively controlled in temperature nor pressure.  The exact imaging of the spectrum onto its detectors is therefore sensitive to changes in ambient air pressure and local temperature; it could also be affected by seismic tremors.  The illumination of UVES is by the telescopic image projected onto the entrance slit, and thus possible variations in stellar image position caused by telescope pointing, atmospheric refraction, or dispersion, may affect the recorded spectra.  Atmospheric conditions and a multitude of instrument settings were obtained from the observation headers available in the ESO archive and correlations with observed values were sought.  

\subsection{Selecting exposures and fitting wavelength scales}

The very first exposures of the observing night were recorded through large airmasses; imperfect seeing conditions did not give the highest S/N ratios (Fig.~\ref{fig:metadata}). Although nothing directly erroneous in their data was seen, as a precaution, the first two exposures were removed from further analysis.  Around each selected spectral line, an interval of $\Delta\lambda$ = 1 nm was extracted for line fitting and continuum placement.  To determine the wavelength positions (and to later place all lines on a common scale), following tests with various functions line profiles were fitted with a five-parameter modified Gaussian function of the type $y_0 + a \cdot \exp[-0.5\cdot(|x-x_0|/b)^c]$.  This function provides values for the central wavelength $x_{\textrm{0}}$, the line depth $a$, the continuum level $y_{\textrm{0}}$, and $b$ as a measure of the line width.

Seemingly clean spectral lines were selected from all the BLUE, REDL, and REDU wavelength regions, and their apparent changes during the night were examined. Among spectral lines in individual exposures during the transit, and immediately afterward, no significant variations could be detected in either fitted line widths or depths, indicating that the spectrometer focus had remained stable throughout the night.  A clear dependence was only seen in the fitted wavelengths.  These values were transformed to first heliocentric and then astrocentric values (Fig.~\ref{fig:metadata}). 

Figure \ref{fig:nominal_wavelengths} shows the trends in observed wavelength shifts of lines in the REDL region during the planet transit, and for some time thereafter, plotted relative to an average value outside transit.  Each dot denotes the fitted wavelength to one specific spectral line.  The drifts in wavelength during the night are gradual and slow (rather than random between successive exposures).  They seem to be common for all spectral lines, i.e., the full spectra seem to shift together; these are not random shifts between different lines.  A typical spread of $\sim$25 m\,s$^{-1}$ is consistent with the photometric noise expected in individual line profiles. 

As seen in Fig.~\ref{fig:nominal_wavelengths}, the UVES wavelength stability is not fully on the level of exoplanet search instrument, and does not permit us to precisely follow radial-velocity changes during the planet transit.  However, the presence of the Rossiter-McLaughlin effect is suggested by these data, superposed on their overall drift pattern.  Since transits are repetitive, the expected signature can be acquired from some other, more precise measurement.  The Rossiter-McLaughlin effect observed by \citet{winnetal05} is overplotted in gray.  Their data from Keck-1 HIRES spectra had a spectral resolution $\lambda$/$\Delta$$\lambda$ $\sim$80,000 (similar to here) and were measured through an iodine absorption cell, thus referring to a limited interval of the spectrum around 500-630 nm, closely overlapping with our REDL region. 

Lines in the shorter wavelength BLUE region generally are more blended and photometrically noisier, but show a very similar dependence that is similar to the REDL lines during transit, and immediately afterward.  There is no systematic change of any fitting parameter, except the wavelength.  The drifts are similar but, toward the end of the night, the BLUE curves start to deviate from the REDL curves.  

The patterns largely mirror the changing airmass during the night (Fig.\ \ref{fig:metadata}).  The housekeeping data indicate that the Atmospheric Dispersion Correction mode was off  and the telescope image derotator was used to place the atmospheric dispersion vector along the spectrometer slit to avoid slit losses.  Possibly, at large airmasses the chromatic illumination of the spectrometer may have changed, causing some slight velocity drift between lines in different spectral regions.  Also, a division of REDL lines into four wavelength groups indicates slight drifts between these groups (however only during the exoplanet transit); the shift between the shorter and longer wavelength parts are on order 15 m\,s$^{-1}$s.  Physical effects causing such types of dependence could originate from the effective planet size being color dependent, but the photometry by different authors is not very conclusive within this limited spectral range \citep{jhaetal00, knutsonetal07} and we do not pursue this any further.

Since the amount of gradual wavelength shifts is consistent with the known stability of the spectrometer, the observed drifts are most likely normal optomechanical drifts in the VLT-UVES system, and one can then place the observed line profiles on a true wavelength scale by shifting it for each observation to coincide with the Rossiter-McLaughlin wavelength position as measured by \citet{winnetal05} and interpolated to the corresponding transit epochs.   Although the very precise velocity amplitudes could still depend on exactly how the correlation calculations of the Keck-1 HIRES iodine-cell measurements compare with our line fitting, we believe that possible differences must be negligible in the present context (given similar spectral resolutions in similar wavelength regions); this is why we adopt these radial velocities for further data handling.  Figure~\ref{fig:fitted_wavelengths} shows the resulting distribution of individual line wavelengths for thus fitted exposures during and immediately after the transit.  

This procedure of combining spectrophotometry from one telescope with radial-velocity data from another could have been avoided if a single spectrometer with superior performance had been available.  Observations of HD~209458 from adequately wavelength-stable instruments (e.g., ESO HARPS) were examined for this program but their photometric noise level was not found sufficient; however, the newest instruments on very large telescopes such as PEPSI \citep{strassmeieretal15} and ESPRESSO \citep{pepeetal14} should be fully adequate.

\begin{figure}[H]
\centering
\includegraphics[width=\hsize]{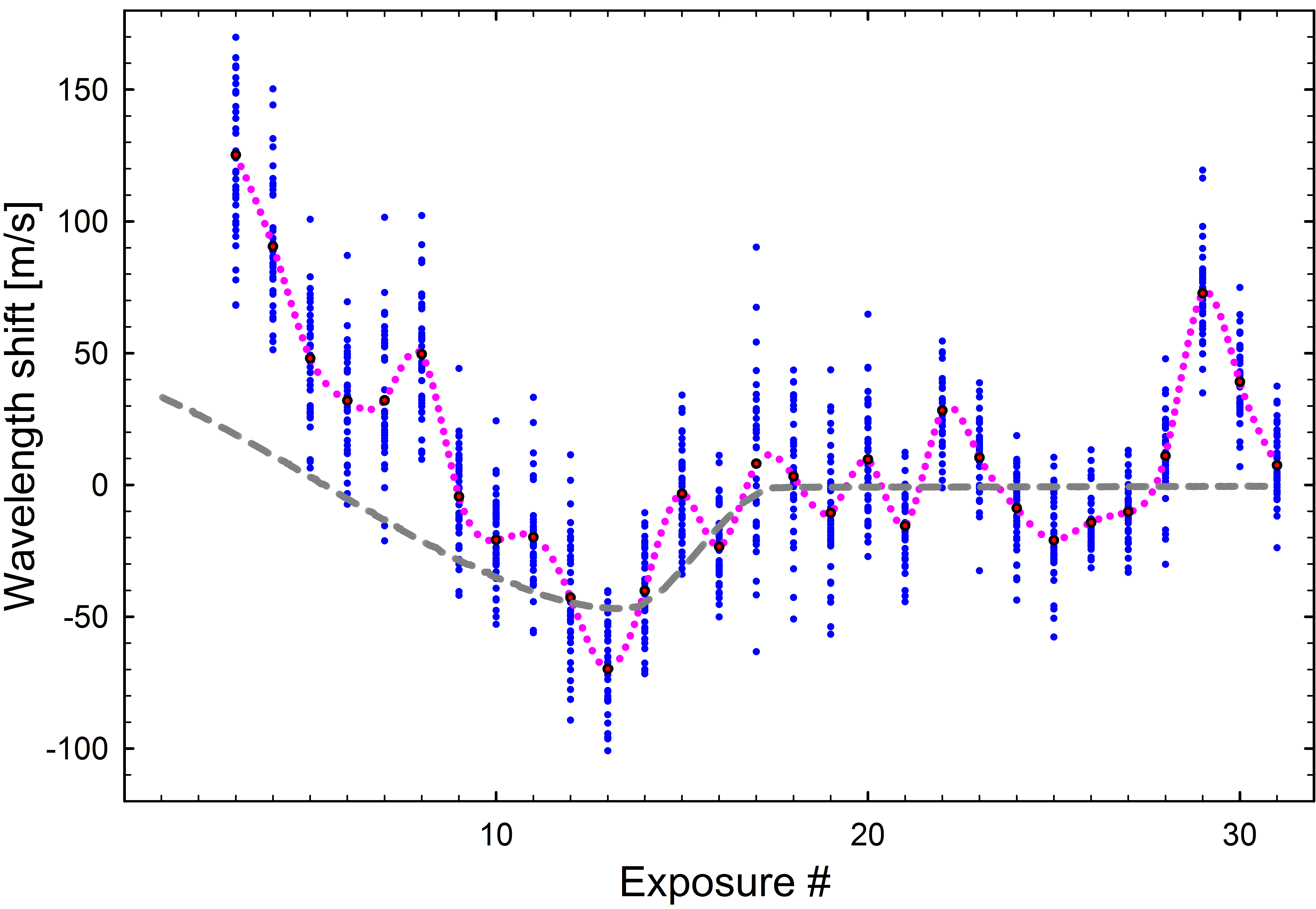}
\caption{Observed wavelength positions of individual lines (dots) drift slowly but systematically between successive exposures. The shifts are common for all spectral lines (trend in dotted red).  Superposed (dashed gray) is the expected Rossiter-McLaughlin effect, as measured with an iodine cell at the Keck-1 HIRES spectrometer, using a similar spectral resolution in the same wavelength region \citep{winnetal05}.  The wavelength scale shows relative shifts only. }
\label{fig:nominal_wavelengths}
\end{figure}

\begin{figure}[H]
\centering
\includegraphics[width=\hsize]{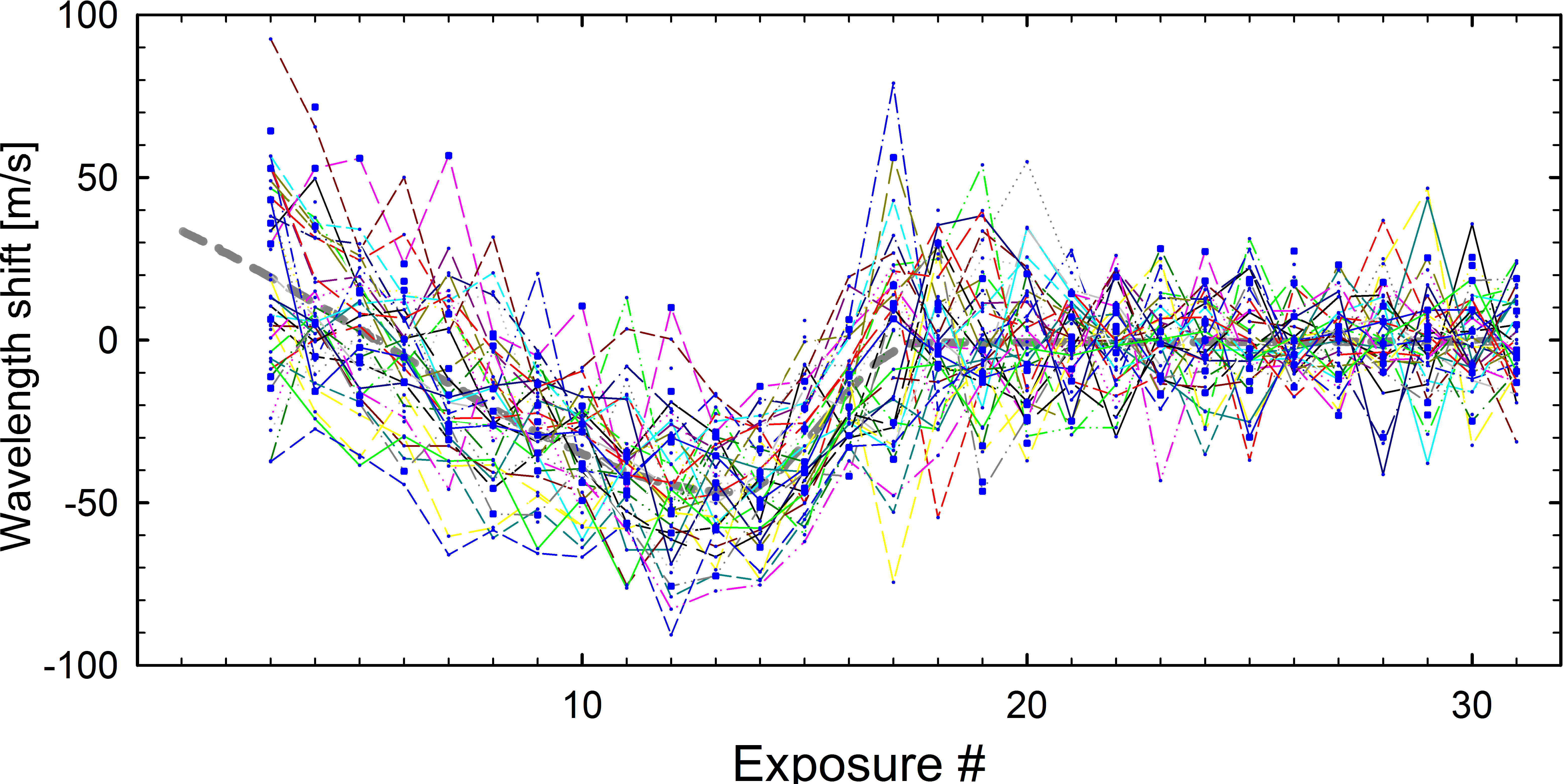}
\caption{Wavelength positions of selected individual Fe I lines with the exposure-averaged wavelengths shifted to those of the measured Rossiter-McLaughlin effect (dashed gray).  The increased noise around exposure \#17 is caused by the S/N then dropping to `only' $\sim$300; cf.\ Fig.~\ref{fig:metadata}. }
\label{fig:fitted_wavelengths}
\end{figure}

\subsection{Reference spectrum from outside transit}

While individual spectral lines recorded during and just after transit do not show any noticeable line profile changes, this does not hold for exposures taken later in the night.  Until exposure \#31 (out of 56), the wavelength calibration for the CCD detectors was based on one particular exposure with a thorium-argon emission-line lamp \citep{hanuschiketal02}, but thereafter no less than seven recalibrations had been made during the rest of the night.

Intensity ratios of spectra from successive exposures show these to be quite stable for spectra reduced with a given set of calibration parameters, but deviant when compared between different calibration setups.  Differences are typically on the order of 0.5\% but may reach 2\% in spectral regions of strong contrast, i.e., across absorption lines.  For ordinary spectroscopy, these effects may in most cases be negligible, e.g., if S/N not much better than $\sim$100 is required.  Presumably, these recalibrations were intended to improve the wavelength precision for the original program, but the photometric stability suffers; this is probably because different CCD pixels with different sensitivity patterns are weighted differently toward retrieved spectral intensities. 

To avoid possible systematics, we discarded all such exposures with deviant calibrations.  For a reference stellar spectrum outside transit, an average of eight exposures was formed; all of these exposures retained the same calibration exposure as during transit, i.e., they did not extend beyond exposure \#31.  The first exposures just outside transit were also not included because of the possibility of an extended exoplanet atmosphere that might show some spectral features, and because at that moment there were some light clouds passing (Fig.\ \ref{fig:metadata}), which might have changed the amount of telluric water-vapor absorption.

\begin{figure}[H]
\centering
\includegraphics[width=\hsize]{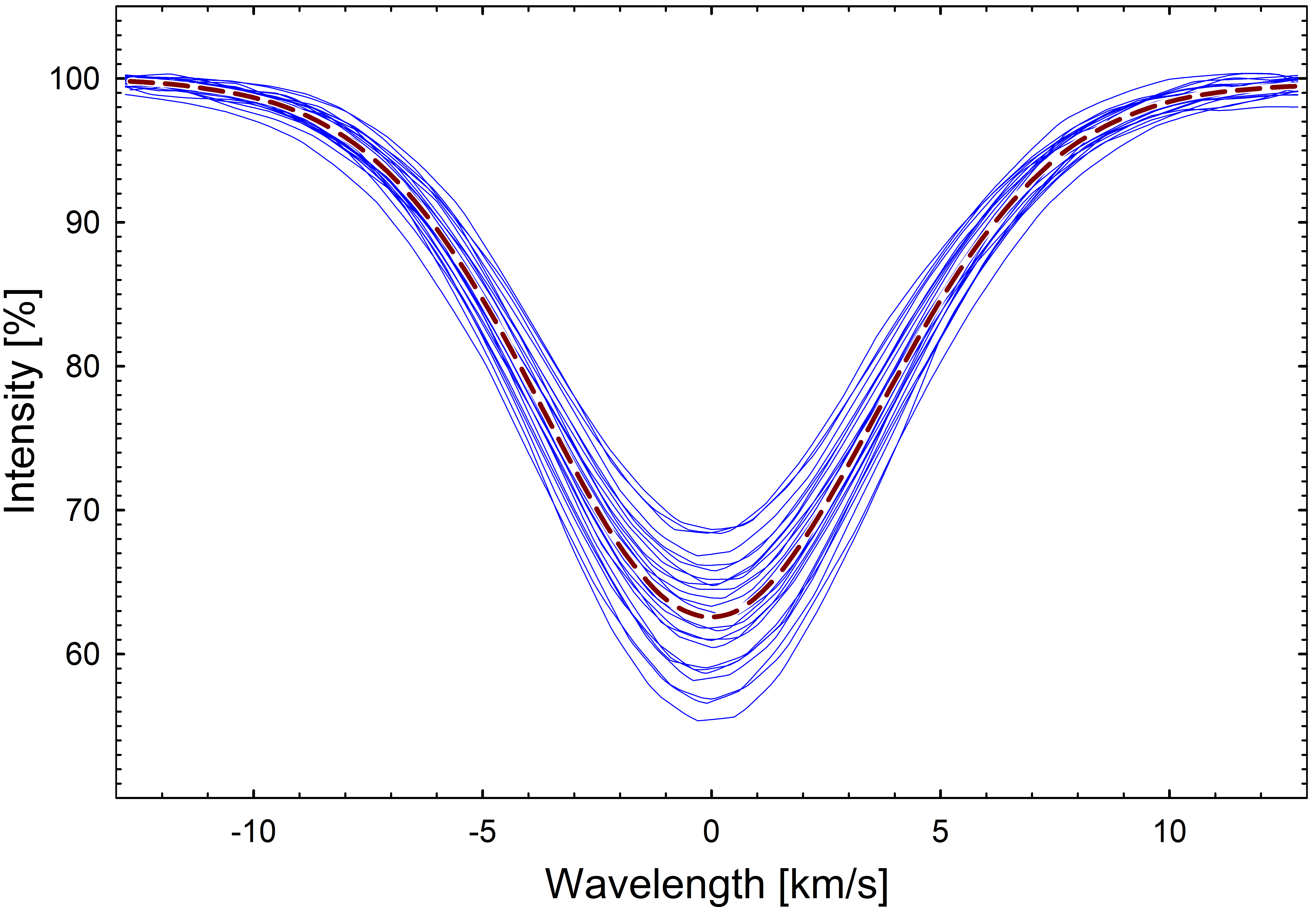}
\caption{Averaging 26 `similar' photospheric Fe~I lines in HD~209458, selected to be largely unblended and of closely similar strengths (Fig.~\ref{fig:line_statistics}).  Their average (dashed) `synthesizes' a representative profile of such a strength in that particular wavelength region.  For this reference profile averaged over several exposures outside transit, the photometric signal-to-noise ratio approaches $\sim$7,000. }
\label{fig:26_fei_lines}
\end{figure}

\subsection{Separating lines into groups}

The distribution of line depths (Fig.~\ref{fig:line_statistics}) permits us to subdivide the sample into subgroups of weaker and stronger lines.  A choice was made to split the groups at a residual observed line depth =~55\,\% of the continuum.  In this finally selected sample, there are 26 weaker lines and 11 stronger lines.  The average wavelengths and excitation potentials are <$\lambda$> = 614 nm, <$\chi$> = 3.3 eV, and 619 nm and 2.7 eV, respectively, for the weaker and stronger line groups.

These lines are next superposed on a common scale.  The continuum intensity value of 100\,\% as well as the line-center position is obtained from the modified five-parameter Gaussian fit, and wavelength values converted to equivalent Doppler velocities.  Figure~\ref{fig:26_fei_lines} shows the group of the 26 weaker lines (from the reference spectrum), together with their average.  This average profile thus represents a typical Fe~I line of such a strength in this wavelength region.  Having started with original exposures with S/N $\sim$500, then averaging over 26 lines, and then over 8 exposures, the nominal signal-to-noise ratio of this average profile approaches $\sim$7,000.

\section{Spatially resolved line profiles}

\subsection{Stellar limb darkening}

Before spatially resolved spectra can be retrieved, the relative amount of flux hidden from the stellar disk area at each transit phase must be known.  The amount of relative flux changes across the disk because of stellar limb darkening, and this amount can be obtained if we know the limb-darkening function and the exoplanet transit geometry.  The numerous photometric transit light curves recorded for HD~209458 confirm that no significant spots seem to be present on this star.  The limb-darkening functions used here are those deduced specifically for HD~209458 by \citet{hayeketal12} from 3\mbox{-}D stellar model atmospheres in passbands including the red SDSS $r'$, whose effective wavelength of 620.4 nm \citep{fukugitaetal96} closely coincides with the average for our spectral-line selections.  However, we stress that our method of spatially resolved line reconstruction does not depend on any theoretical predictions of limb darkening; only the product of planetary area and stellar local brightness is required and such data could be taken directly from observed transit photometry.  Separating the values for the planet area and limb darkening makes it easier to discuss error budgets, and a limb-darkening model enables us to extrapolate continuum intensities to the stellar disk center that is not sampled during the exoplanet transit.  

\begin{figure}[H]
\centering
\includegraphics[width=\hsize]{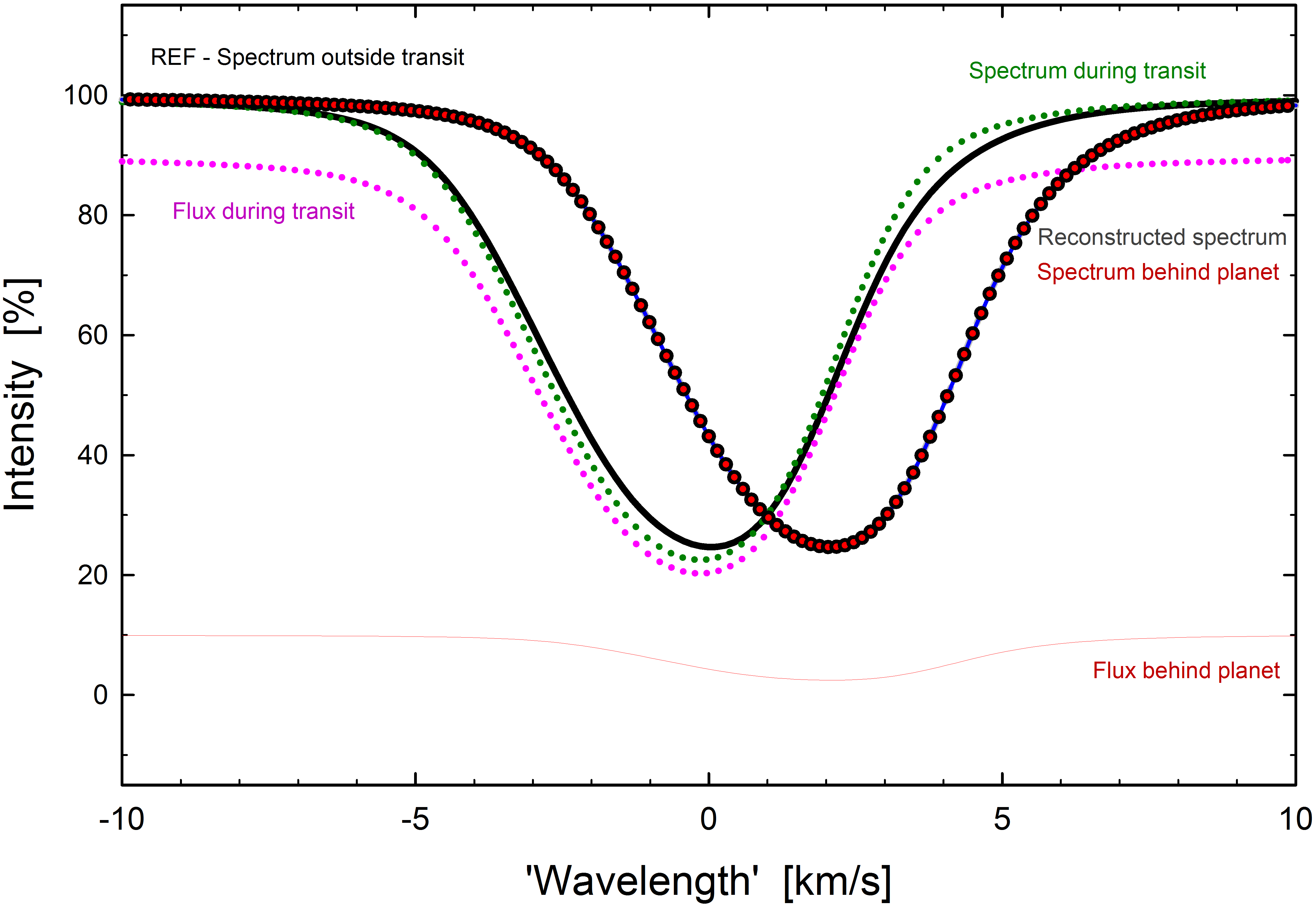}
\caption{Different spectral-line components treated in the reconstruction of spatially resolved spectra.  The spectrum behind the planet is obtained as that line profile (weighted with the amount of flux temporarily obscured by the planet) that -- summed with the temporarily observed line profile -- produces the stellar reference profile outside of transit.  For clarity, the planetary signal here is greatly exaggerated.}
\label{fig:principles}
\end{figure}

\subsection{Reconstructing spatially resolved profiles}

The principle for profile reconstruction is shown in Fig.\ \ref{fig:principles}.  The spectroscopically observed profiles are different from each transit phase and from the reference profile from outside transit. The diminution of the flux at each transit phase is taken from limb-darkening functions and planet-size determinations fitted to independent photometric observations.  The spectrum behind the planet is obtained as that line profile, weighted with the appropriate fractional flux, that, together with the then observed line profile, produces the profile without a planet outside transit.  The resulting spectral continuum level can be expressed relative to the intensity at stellar disk center or to the full-disk average (thus showing effects of limb darkening) or it can be normalized to a local intensity of 100\,\% in the more common spectral format. In this illustration, the impact of the planet is exaggerated by an order of magnitude: even large planets do not cover much more than 1\% of the surface of solar-type stars.

\begin{figure}[H]
\centering
\includegraphics[width=\hsize]{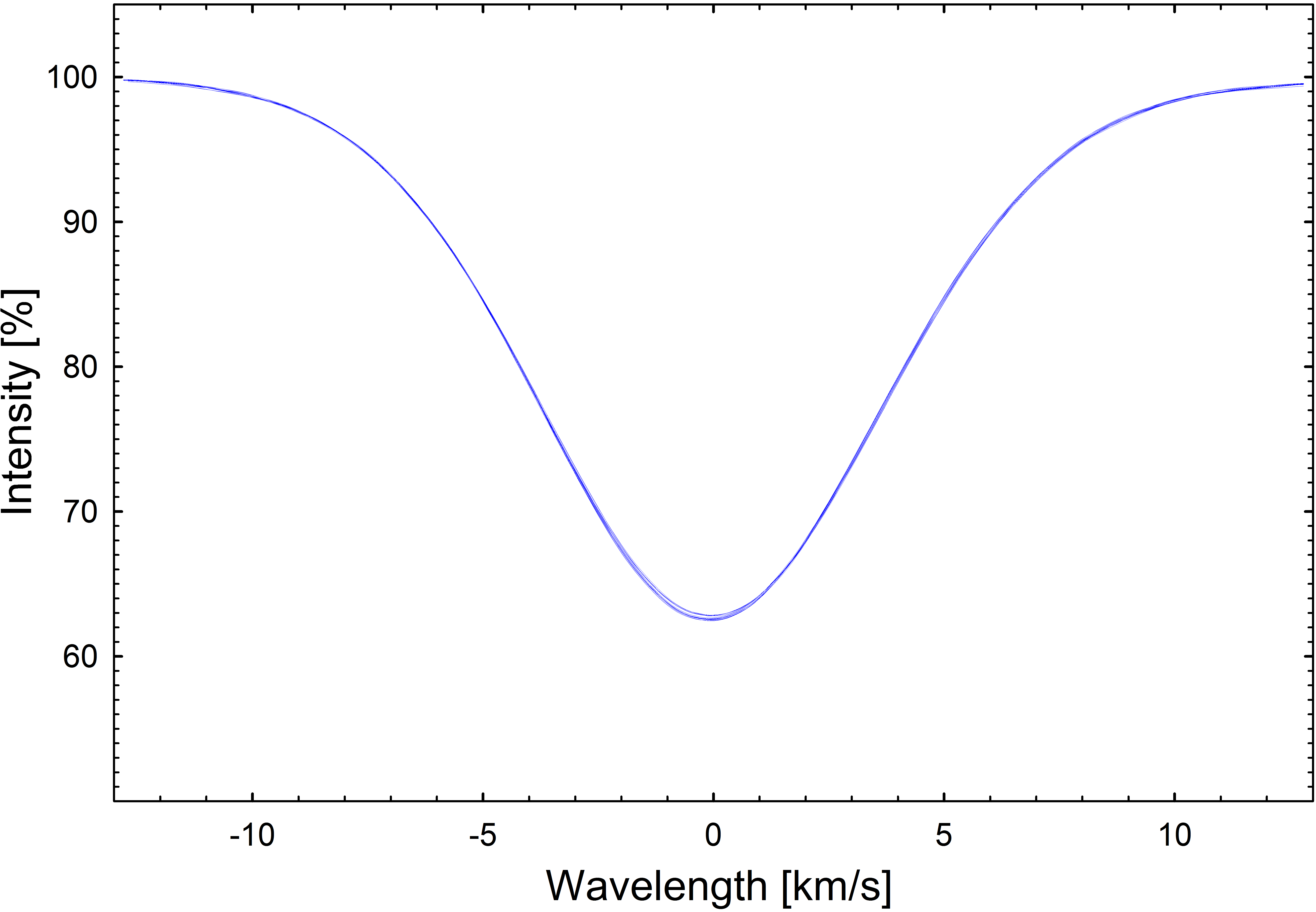}
\caption{Representative profiles of the weaker Fe~I line and its gradual changes.  Each profile, which is the average of 26 Fe~I lines, is plotted for 14 successive exposures during the exoplanet transit, showing the degree of uniformity reached.  To appreciate the slight changes during the transit, this plot should be viewed highly magnified; corresponding line ratios are in Fig.~\ref{fig:weak_line_ratios}. The photometric signal-to-noise ratio here is $\sim$2,500.}
\label{fig:weak_line_14_exposures}
\end{figure}

\subsection{Transit sequence of averaged Fe~I profiles}
    
Figure~\ref{fig:weak_line_14_exposures} shows an overplot of the averaged weaker Fe~I line for 14 successive exposures during the Osiris transit.  Compared to the schematics of Fig.~\ref{fig:principles}, the spectral-line variations in the actual data are tiny.  Here, the averaging over 26 different lines produces a photometric S/N $\sim$2,500 that begins to be useful when combined with the still lower noise reference profile from outside transit.  While this plot shows the degree of uniformity reached, the differences between successive exposures are too subtle to be easily appreciated. These are better seen in Fig.~\ref{fig:weak_line_ratios}, which shows the successive ratios of each line profile to the reference profile, in the same format at the theoretical curves in Figs.~9 and 10 of Paper~I.

\subsection{Reconstructed line profiles}

To compute the flux temporarily hidden during transit, the projected planet area was taken as 1.5\,\% of the stellar disk, with limb darkening from \citet{hayeketal12} for the passband SDSS $r'$; these values are deduced from photometry.  The center-to-limb position of the planet from the transit trajectory was computed from the impact parameter =~0.51, as deduced by various authors from the measured Rossiter-McLaughlin effect.

\begin{figure}[H]
\centering
\includegraphics[width=74mm]{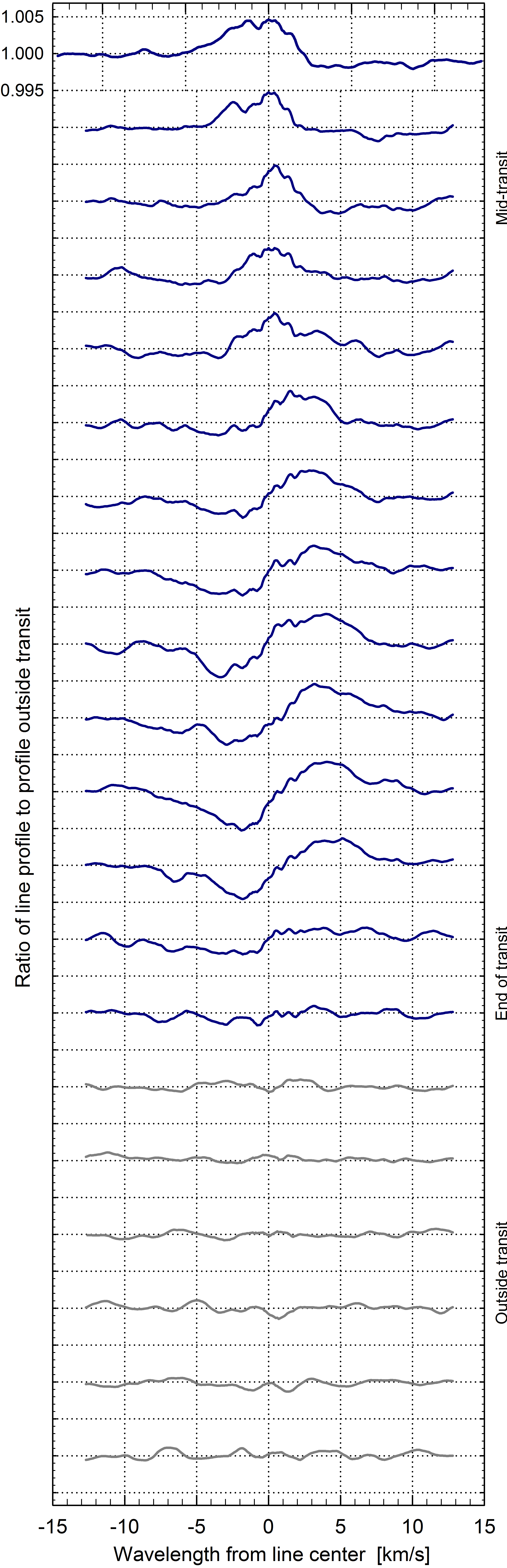}
\caption{Ratios, during successive exposures, of observed line profiles to the reference profile from outside transit.  The sequence starts with the planet already in transit and extends after it.  Time increases from top down.  Line profiles are averages of the group of 26 weaker Fe~I lines in HD~209458. }
\label{fig:weak_line_ratios}
\end{figure}

Figure~\ref{fig:reconstructed_26_fei_lines} shows the reconstruction of a sequence of profiles from 11 epochs during the transit.  The final epochs are not shown as they become quite noisy, both due to a decreased signal caused by limb darkening and from the planet leaving the stellar disk (Fig.~\ref{fig:transit_geometry}).  Already a number of features are becoming visible, in particular that because spatially resolved lines are not subject to rotational broadening, they are substantially deeper and narrower than those from the spatially averaged disk.  However, it is clear that, even at original S/N $\sim$\,2,500, reconstructed lines are still rather noisy (as indeed expected from the simulations in Paper~I).  An averaging over groups of several reconstructed lines somewhat decreases the noise (at a certain cost in spatial resolution) and makes the spatially resolved spectra easier to appreciate; see Fig.~\ref{fig:reconstructed_averaged_limb}.  There the planet positions are shown to scale on the stellar disk, although this averaging of spectra means that the signal comes from a somewhat smeared-out segment along the transit path.  By fitting functions to the reconstructed profiles, overall trends become easier to discern.  The profiles from 14 transit epochs, fitted to modified Gaussians, are shown in Fig.~\ref{fig:profile_fits_14_epochs}.  

Analogous analyses were made for the group of 11 stronger lines with quite similar results.  Their smaller sample makes the data somewhat noisier (cf.\ Fig.~11 in Paper~I), although this is partially remedied by the greater number of wavelength points across these broader lines.

\begin{figure}[H]
\centering
\includegraphics[width=\hsize]{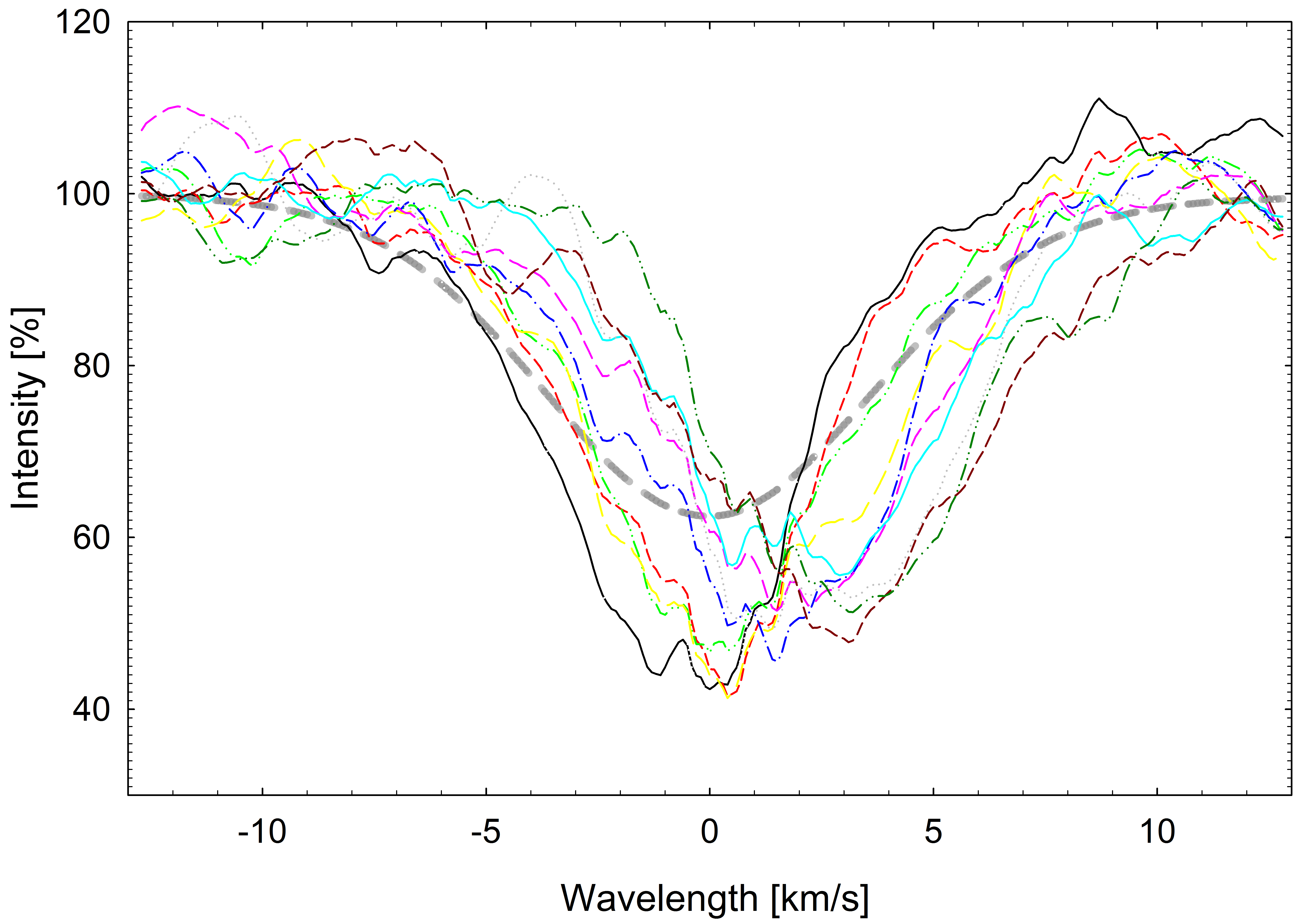}
\caption{Reconstructed profiles for the weaker Fe~I line in a sequence of 11 spatial locations across the disk of HD~209458.   Spatially resolved lines are not subject to rotational broadening and are substantially deeper than in the disk-averaged spectrum outside transit (dashed gray).  During transit, the profiles shift toward longer wavelengths, illustrating both the stellar rotation at the latitude of transit and the prograde orbital motion of the exoplanet.  The intensity scale is the local scale at each disk position. }
\label{fig:reconstructed_26_fei_lines}
\end{figure}

To better display the changing line shapes during transit, these line shapes were centered on the same wavelength removing the Doppler shift induced by stellar rotation; averages over several adjacent exposures were formed from five exposures closer to disk center, where line changes should be modest, to just two exposures near the limb, where line changes are expected to be rapid.  The same treatment was applied to both the weaker and stronger line groups; these fitted profiles are in Fig.~\ref{fig:profile_fits_weak_strong} for three different average center-to-limb positions.  This clearly shows how both groups of lines become systematically broader (e.g., in the half-width sense) and shallower when going from disk center toward the limb, which is a signature predicted from 3\mbox{-}D hydrodynamic models, as discussed in Paper~I.  For a more precise quantitative evaluation, however, the line broadening induced by the finite spectrometer resolution has to be considered.

\subsection{Stellar rotation}

The stellar rotation is known previously from both measured line broadening and from the Rossiter-McLaughlin effect, but can be verified again.  During transit, the reconstructed line profiles systematically shift toward longer wavelengths, reflecting both the stellar rotation velocity vector at each particular planet position and the prograde orbital motion of the exoplanet.  The apparent latitude of transit has been determined to $27^{\circ}$ and the tilt of the stellar rotational axis to some small value around $4^{\circ}$ \citep{winnetal05}.  The velocity shift observed during our partially covered transit of $\sim$3 km\,s$^{-1}$ is fully consistent with these other determinations indicating $V$\,sin$i$~$\sim$\,4 km\,s$^{-1}$ but would need improved data to be much better constrained.

\begin{figure}[H]
\centering
\includegraphics[width=\hsize]{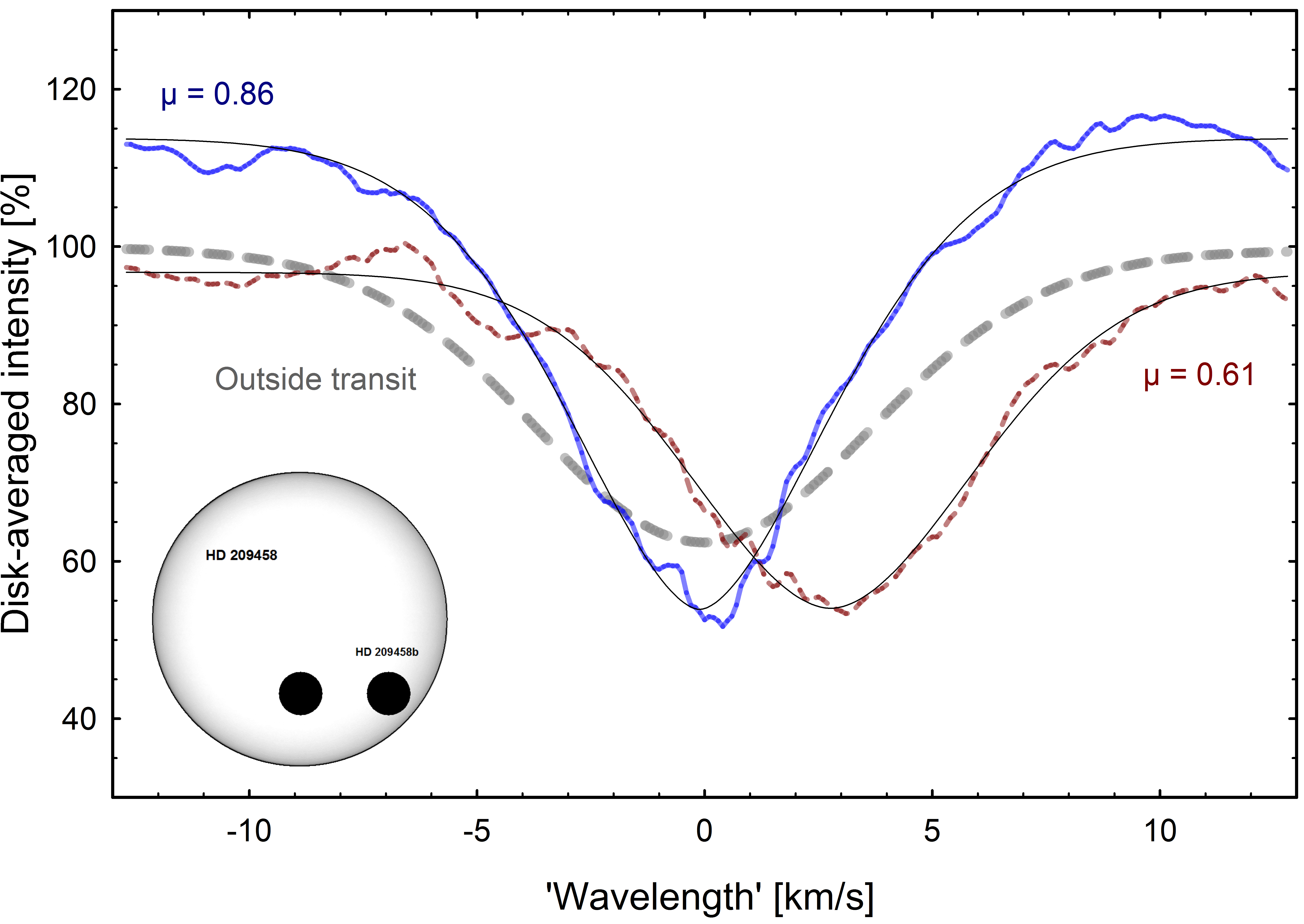}
\caption{Center-to-limb variations in reconstructed line profiles. The solid blue curve is the average of five exposures near stellar disk center with mean $\mu$ = 0.86; the dashed dark red curve is the average over four exposures closer to limb, <$\mu$> = 0.61.  The fitted thin curves are modified Gaussians;  the observed profile outside transit is bold dashed gray.  The intensity scale is that of the stellar disk-averaged flux.  Because of limb darkening, the local intensity close to disk center is higher, and that further away is lower than this average value.  The planet size and disk positions are to scale. }
\label{fig:reconstructed_averaged_limb}
\end{figure}

\subsection{Chromospheric lines}

Some efforts were made in retrieving line profiles of also very strong lines from the upper photosphere or of a more chromospheric character, including Ca~II H \& K, H\,$\beta$, H\,$\gamma,$ and H\,$\delta$ in the UVES BLUE spectral region, Na I D$_{\textrm{1}}$, D$_{\textrm{2}}$, and H\,$\alpha$ in REDL and the infrared Ca~II triplet in the REDU segment.

Such lines are few in number, mutually different, and thus the option of averaging multiple similar lines is not applicable.  On the other hand, their much greater wavelength widths (as compared to photospheric Fe~I) permit a certain smoothing to decrease the photometric noise (at a cost in spectral resolution).  Further aspects (both problems and possibilities) are that the stellar lines may appear in emission inside or outside the stellar limb, and the same lines could also be present in the extended atmosphere around the exoplanet.  These strong lines do not primarily obtain their widths from photospheric motions and any detailed comparisons to 3\mbox{-}D models will require spectral-line synthesis in such upper-atmospheric conditions as well.  Although their interpretation could thus be complex, their examination provides some guidance toward the requirements for higher fidelity observations in the future.  Examples of attempted reconstructions were given by \citet{dravinsetal15}.

\begin{figure}[H]
\centering
\includegraphics[width=\hsize]{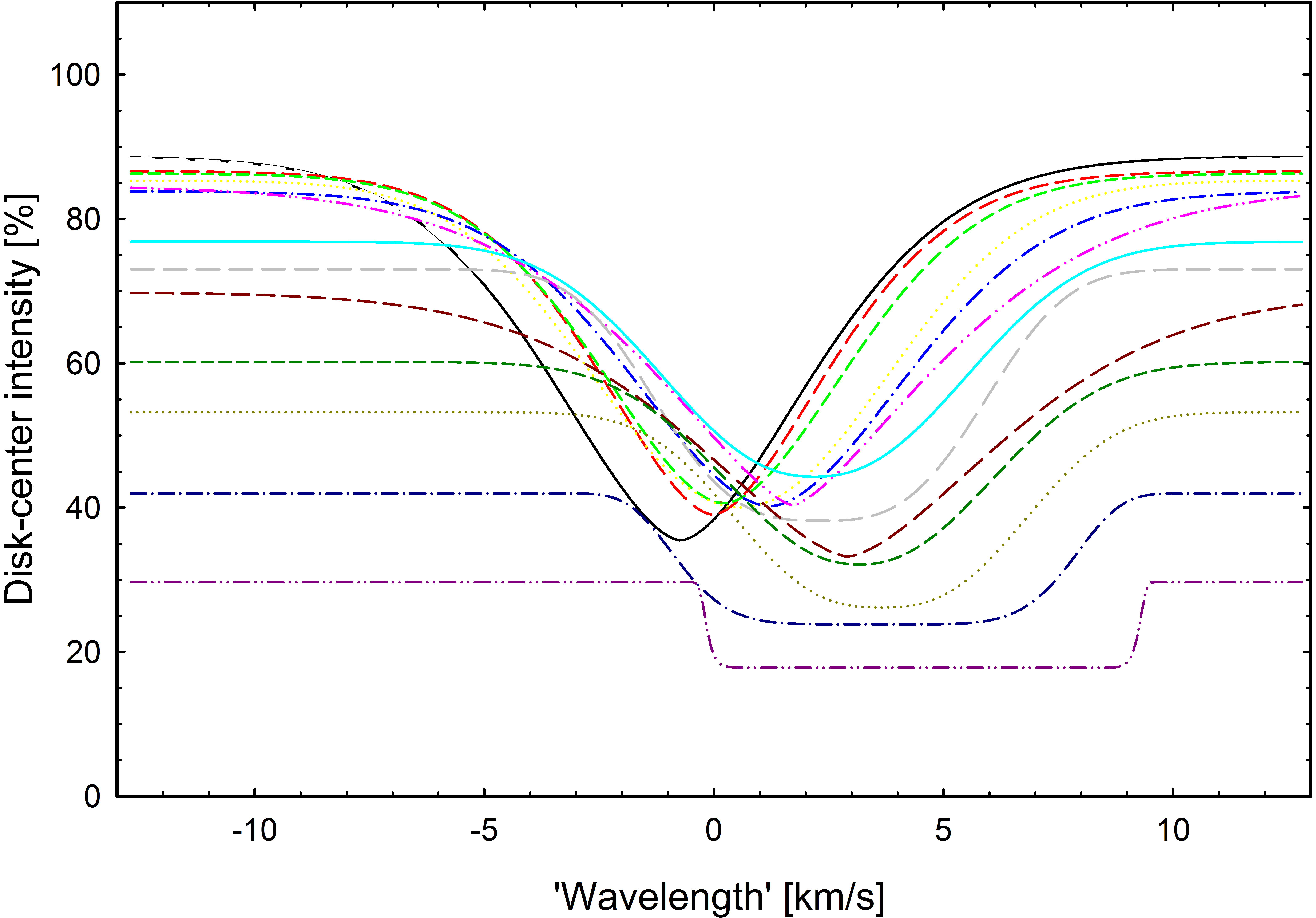}
\caption{Fitted line profiles at 14 observed epochs during exoplanet transit.  The intensity scale is here normalized to stellar disk center, thus also showing the limb darkening at the successive planet positions.  Data very close to the limb suggest quite broad profiles, although these measurements are somewhat uncertain. }
\label{fig:profile_fits_14_epochs}
\end{figure}

\begin{figure}[H]
\centering
\includegraphics[width=\hsize]{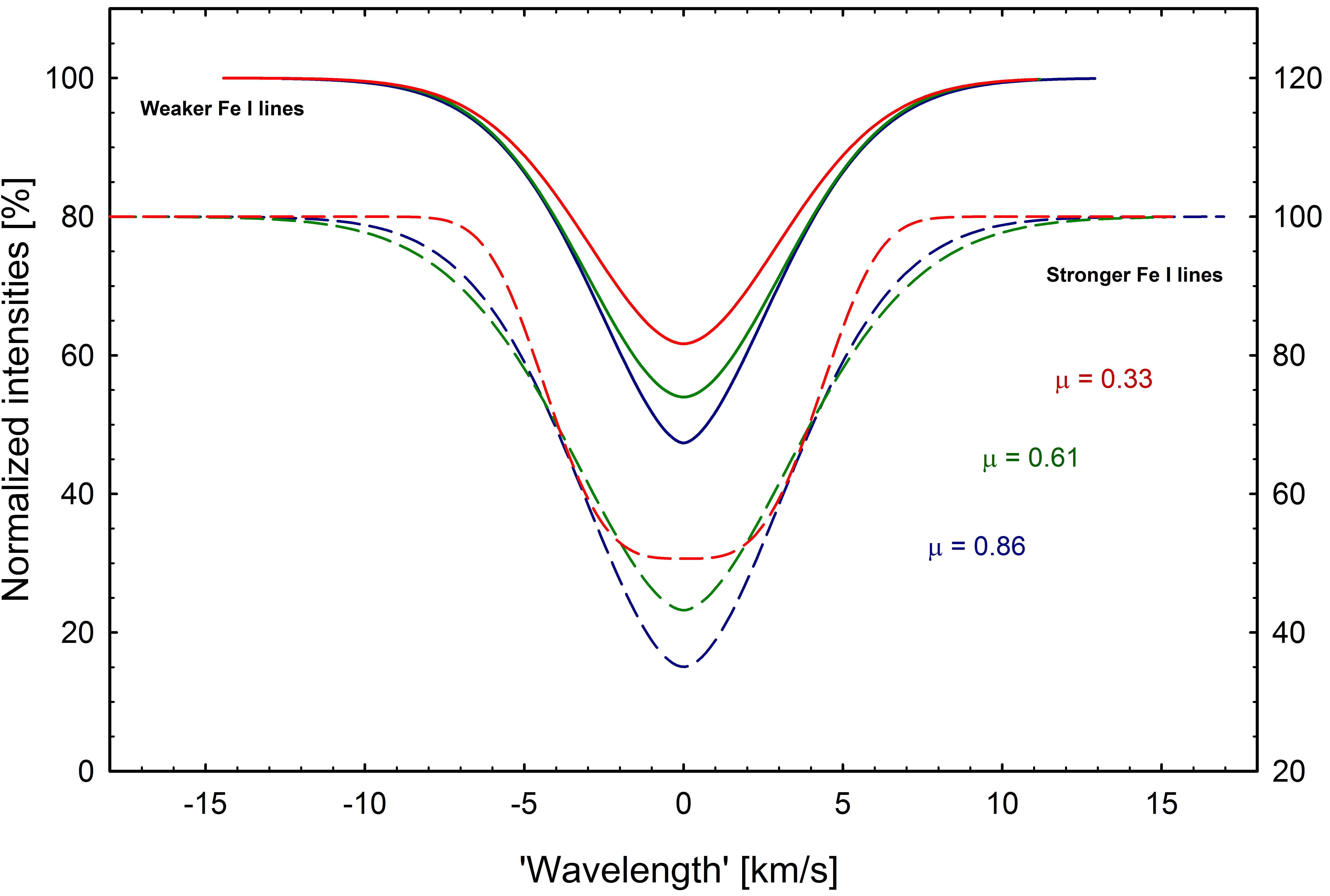}
\caption{Sequence of Fe~I line profiles for the group of 26 weaker lines (left intensity axis) and 11 stronger lines.  For the disk positions <$\mu$> = 0.86, 0.61, and 0.33, the spatially reconstructed profiles are averages over 5, 3, and 2 exposures.  The fitted modified Gaussian functions are centered and normalized for easier comparison. }
\label{fig:profile_fits_weak_strong}
\end{figure}

An additional observational issue for these lines, whose widths might be not negligible compared to the blaze efficiency function in echelle spectrometers, is the precise fitting of the spectral continuum. This is an issue because, on our required accuracy levels, the varying airmass and atmospheric absorption between successive exposures affect the slope of the observed spectral continuum.  Further, especially in regions of the longest red wavelengths, artifacts may originate from the varying strength of telluric water vapor lines during, and outside, the hours of transit.  The reconstruction yields that line profile, whose intensity-weighted summation with the observed transit profile, equals that from outside transit.  Any change of telluric line strengths between different transit epochs causes the deduced spectrum from behind the exoplanet to carry all this change.  Since it has only a very small weight (corresponding to the small area coverage of the planet), this spectrum must then be significantly modified to account for it all.  An example of how changing telluric absorption in the wings of H$\alpha$, such as that shown in Fig.~1 of \citet{reinersetal16}, causes a spurious `emission' in the reconstructed profile, is in Fig.~9 of  \citet{dravinsetal15}.

\subsection{Systematic noise sources} 

Several calculations were made to examine the effects of plausible random and systematic errors.  One parameter that turns out to carry only a small effect is that of the exact projected planet area.  The uncertainties in planetary area are small anyway, although the value for a gaseous planet may depend on the exact wavelength region.  Modest changes in this area do not significantly change the deduced line profiles, which should permit the averaging of line profiles from more extended wavelength regions as well.

A more critical parameter is the exact wavelength scale.  Reconstructed lines are obtained as subtle differences between profiles at different epochs of exposure, and wavelength errors between those of already $\sim$10 m\,s$^{-1}$ cause disturbances in the reconstructed profiles, shifting the profiles by much more than such errors.  Thus, accurate wavelength control in the observations is required to at least such a level.  For the present work, the UVES original wavelength scale was not adequate; this is why its photometrically good spectra were combined with more precise radial-velocity data from the HIRES iodine cell on Keck-1.  However, instruments with planetary-search capabilities such as HARPS, PEPSI, or ESPRESSO have adequate wavelength stability, although the consistencies of their absolute wavelength scales between separate spectral regions may have to be verified.

\begin{figure*}
\sidecaption
  \includegraphics[width=12cm]{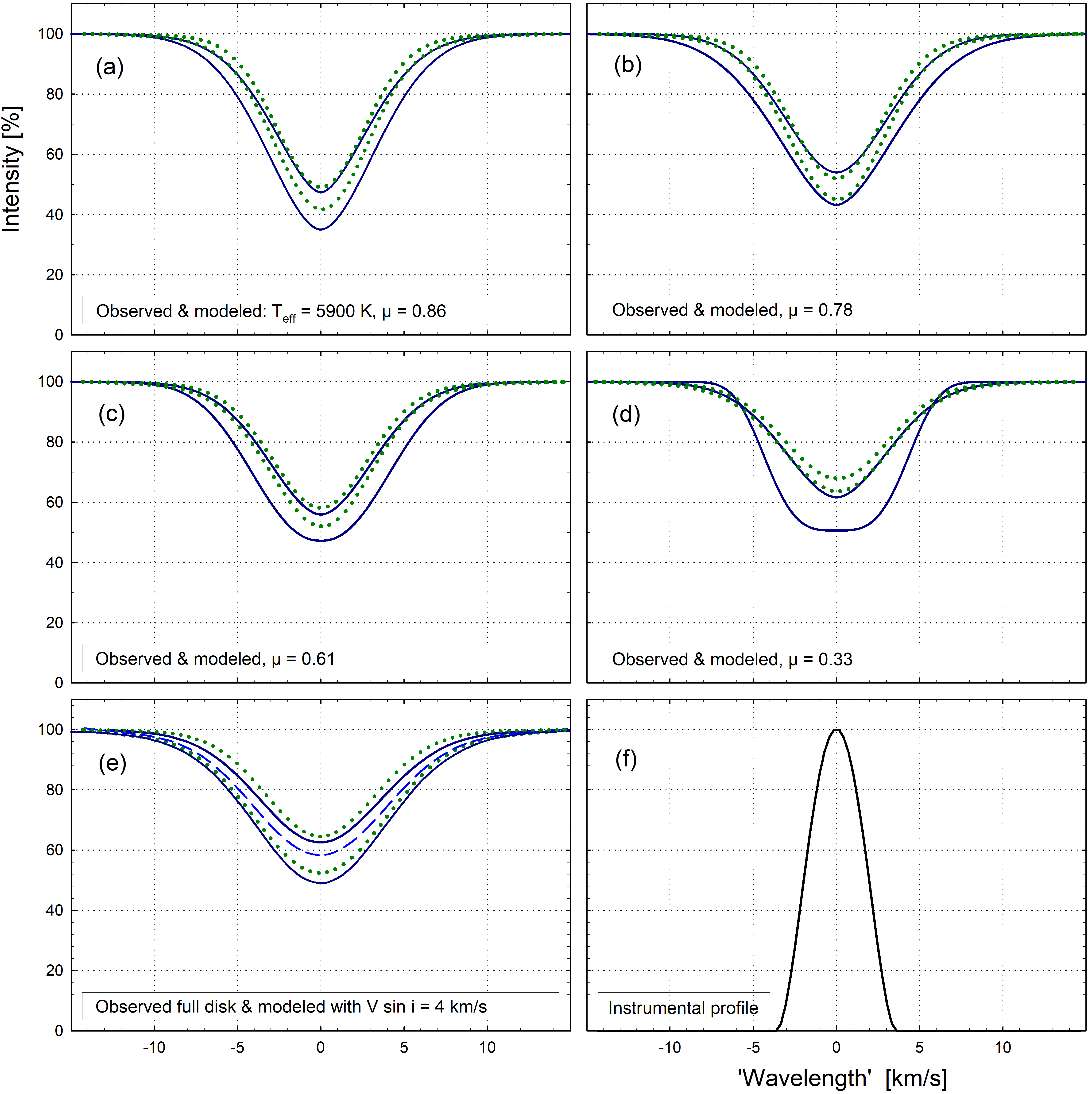}
     \caption{Observed and modeled line profiles for spatially resolved positions across the disk of HD~209458.  (a) Observed profiles for stronger and weaker Fe~I lines at average disk position $\mu$ = 0.86 (solid blue), compared to synthetic profiles ($\lambda$ = 620 nm, $\chi$ = 3 eV) for differently strong lines at the same $\mu$ from a model of T$_{\textrm{eff}}$ = 5900~K (dotted green).  Same types of data in (b) for $\mu$ = 0.78, (c) for $\mu$ = 0.61, and (d) for $\mu$ = 0.33 are shown.  Synthetic full-disk profiles are shown in (e), broadened for stellar rotation of 4 km\,s$^{-1}$, for the T$_{\textrm{eff}}$ = 5900~K model (dotted green) and for 6250~K (dashed blue).  A comparison to observed profiles outside transit (solid) confirms a stellar rotational velocity around such a value.  All synthetic lines are here convolved with the spectrometer instrumental profile in (f) of FWHM = 4.0 km\,s$^{-1}$. }
     \label{fig:instrumental}
\end{figure*}

\section{Comparing to models}

In Paper~I, synthetic Fe~I lines were computed from a grid of CO\,$^5$BOLD models \citep{freytagetal12} for different main-sequence stars in the temperature range between T$_{\textrm{eff}}$ = 3960 and 6730~K.  In particular, this included models for 5900 and 6250~K with surface gravities log~$\varg$ [cgs] = 4.5 and solar metallicity.  Those model parameters bracket the deduced values for HD~209458 of T$_{\textrm{eff}}$ = 6071 $\pm$20 K, log~$\varg$ [cgs] = 4.38, and [Fe/H] = 0 \citep{delburgoallende16}, and synthetic line profiles computed for $\lambda$ = 620~nm and $\chi$ = 3~eV can directly be compared to our spatially reconstructed line profiles with closely similar wavelengths and excitation potentials; see Figs.~\ref{fig:instrumental} and \ref{fig:center-to-limb}.  The observed profiles are fits to the data averaged over adjacent exposures as in Fig.~\ref{fig:profile_fits_weak_strong}.  Their noise level increases from disk center toward the limb, both because of fewer averaged points and because the stellar signal weakens because of limb darkening.  The smaller sample making up the stronger line make its noise level worse than for the weaker line and its profile close to limb should be viewed as uncertain.  It is awkward to specify precise error bars: while the formal fitting errors are small, systematic effects are difficult to quantify.  The full-disk data are the observed reference profiles from outside transit.  Synthetic lines were computed for five different line strengths; the two closest to the observed line strengths are shown; they are interpolated to the same $\mu$-values as those of the observed profiles.

Limitations in how rigorous comparisons can be carried out are set by not only the residual noise levels, but also by the imperfectly known spectrometer response function.  The nominal resolution for the actual setup is $\lambda/\Delta \lambda\sim$80,000, i.e., $\sim$3.8 km\,s$^{-1}$ \citep{dekkeretal00, dodoricoetal00}.  This value refers to a measure such as the width at half maximum, however the precise shape of the instrumental profile is not known.  The synthetic stellar profiles were computed with much higher resolution and thus have to be appropriately degraded for comparison.  A slight additional spectral broadening arises from the spatial smearing across the stellar disk introduced by the averaging of groups of exposures.  In the absence of a precisely known instrumental profile, some credible function must be chosen for convolving the synthetic profiles for comparisons to observed profiles.  \citet{robertsonetal13} reviewed various types of spectrometer line-spreading functions, highlighting uncertainties between the use of different definitions.  After various tests, we chose to use the bisquare function in Fig.~\ref{fig:instrumental}(f) with full width at half maximum of 4.0 km\,s$^{-1}$. 

Two classes of line parameters that can be tested with already the current observations are the center-to-limb behavior of line strengths and widths.  Figure~\ref{fig:center-to-limb} compares theoretically predicted values from models at 5900 and 6250~K to current observations.  Line-bottom intensities are in units of the local spectral continuum at each $\mu$, while the line widths are obtained from five-parameter fits to modified Gaussian functions of the type $y_0 + a \cdot \exp[-0.5\cdot(|x-x_0|/b)^c]$, as in Sect.~5.1, yielding $b$ as a measure of the line width.  The same type of fitting was applied to both observed profiles, as to synthetic profiles (with and without instrumental broadening).  Models predict photospheric lines to become shallower and broader when going from stellar disk center toward the limb; observed profiles by and large follow this trend, but perhaps with a tendency for the observed slope of the line-depth dependence to be shallower than that of the models.

\begin{figure*}
\sidecaption
  \includegraphics[width=12cm]{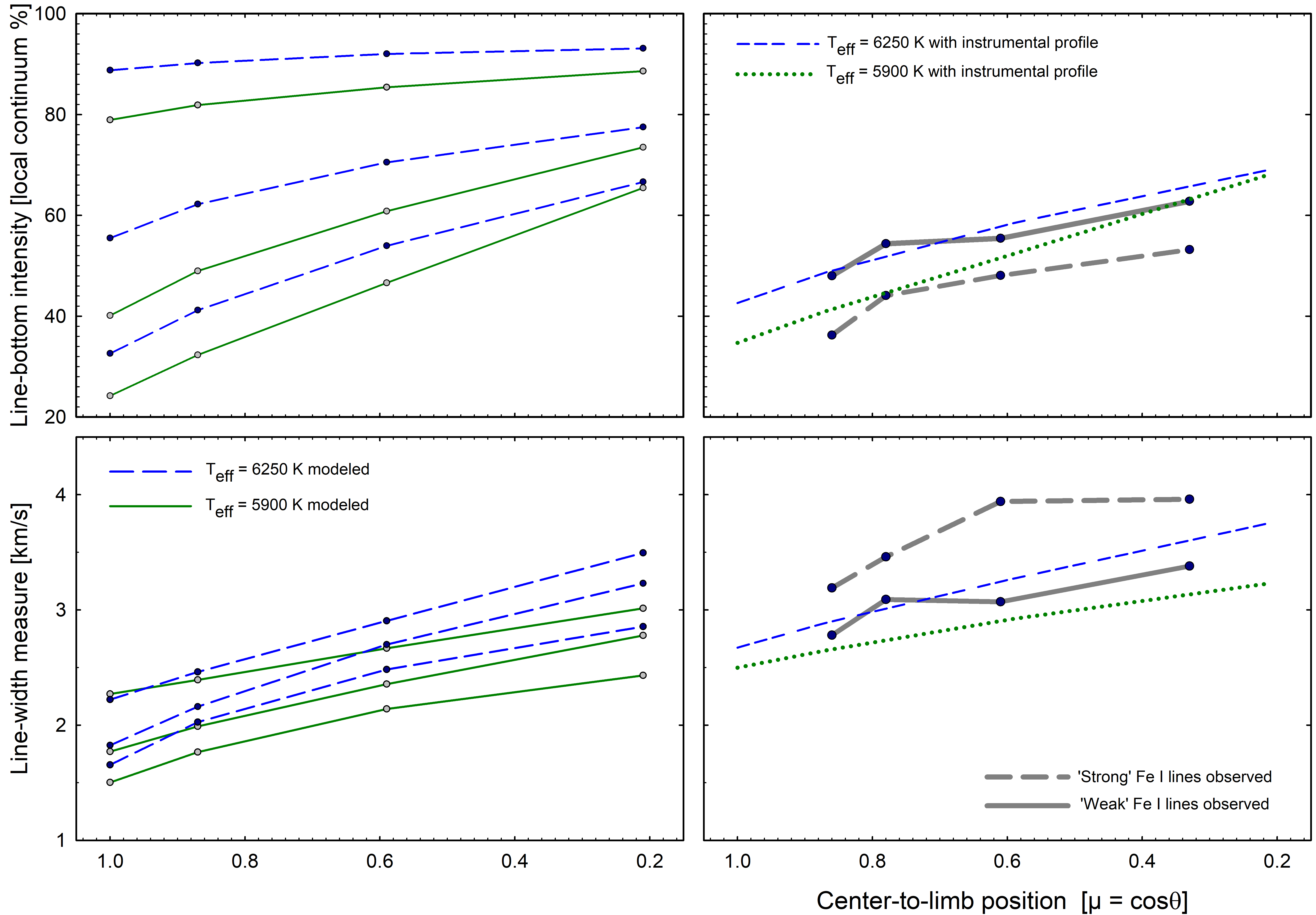}
     \caption{Observed and modeled center-to-limb behavior for line depths and widths.  Predictions from CO\,$^5$BOLD models with parameters bracketing those of HD~209458 are shown.  Left panels show theoretical center-to-limb changes for differently strong lines at full spectral resolution; the right-hand panels compare observations (bold curves) to synthetic profiles degraded with the spectrometer instrumental profile.  Line widths and depths were obtained by fitting modified Gaussian functions to both observed and modeled line profiles. }
     \label{fig:center-to-limb}
\end{figure*}

These comparisons to synthetic lines demonstrate the potential of such observational confrontations with 3\mbox{-}D models, but do not constitute any closed loop of any more complete model adjustments.  The line strengths in the models were not computed to fit any specific observations, in particular lines from the hotter of these two models are generally weaker than those that were feasible to reconstruct from the current data.   Also, retrieving signatures, such as intrinsic line asymmetries, convective wavelength shifts, profiles for weaker lines, or the changes between different excitation or ionization levels, will require better spectral resolution and improved signal to noise. 

Figure~\ref{fig:instrumental} also shows the full-disk and thus rotationally broadened profiles.  Synthetic profiles were computed for different $V$\,sin\,$i$ following \citet{ludwig07} and then convolved with the instrumental profile.  Profiles for 4 km\,s$^{-1}$ from both T$_{\textrm{eff}}$ = 5900 K models (dashed) and 6250 K (dotted) show a close agreement with observed lines from outside transit.  This again confirms the stellar rotational velocity of around 4 km\,s$^{-1}$; however, a more precise determination would require both higher spectral resolution and also a more precise knowledge of the instrumental broadening.

\section{Conclusions and future potential}

As demonstrated above, retrieval of spectral lines across stellar disks is already feasible with present facilities and for current stellar targets.  Although still perhaps somewhat noisy, this -- as far as we are aware -- represents the first case of high-resolution spectra obtained from precisely selected small areas across stellar surfaces.  Moreover, this method is likely to become much more applicable already in the near future.

Here, it should be mentioned that some other means also exist to extract information about stellar spectra and granulation properties behind transiting planets, in particular by measuring the Rossiter-McLaughlin effect or the changing stellar flux in different narrow wavelength bands \citep{ceglaetal16, chiavassaetal17, czeslaetal15, yanetal17}.  However, those methods do not fully address the combined 3\mbox{-}D hydrodynamic modeling together with spectral-line synthesis.

One limitation of the current method is set by the photometric noise that can be reached with existing telescopes and instruments.  The data used here originated from observations with the ESO 8.2 m VLT Kueyen telescope with its UVES spectrometer entrance slit opened up to 0.5 arcseconds to maximize photometric precision at the expense of spectral resolution.  Such a compromise was dictated by the limited brightness of the target star of visual magnitude m$_{\textrm{v}}$ = 7.65.  Other transit sequences recorded with the same instrument at its full spectral resolution were examined, but found to be photometrically too noisy.  Although brighter host stars with large transiting planets are currently not known, numerous surveys, both from the ground and from space, are monitoring brighter stars for possible exoplanet transits; the only other host star of similar brightness, HD~189733A (K1~V), is the subject of another ongoing study.  Given the known statistics of exoplanet occurrence, it is highly likely that suitable transiting planets will soon be found around brighter hosts as well.  Once a target star of visual magnitude m$_{\textrm{v}}$ = 5, say, is found, that will improve the signal by an order of magnitude.  However, the number of sufficiently clean spectral lines available for averaging will depend on stellar parameters such as temperature and speed of rotation. 

A limitation comes from the telescope plus spectrometer combination.  However, even for the current target, and with present instrumentation, an optimized observing program could significantly increase the number of measurable lines through an optimal choice of spectral intervals to be observed (Fig.~\ref{fig:line_statistics}). The Fe~I and Fe~II line groups could be expanded by adding photospheric lines from other metals, such as Ti~I and Ti~II, which carry similar signatures of atmospheric hydrodynamics \citep{dravins08}.  The signal would improve if one or multiple transits were completely covered from start to end and if spectrometer calibrations were kept constant. 

To fully retrieve atmospheric signatures in not only overall line profiles but also in their asymmetries and wavelength shifts, very high spectral resolution has to be combined with excellent wavelength stability; these conditions were not available for the current study.  However, promising developments are in progress \citep{dravins10}.  The PEPSI spectrometer \citep{strassmeieretal15} at the Large Binocular Telescope combines extended spectral coverage with resolutions $\lambda$/$\Delta\lambda$ up to $\sim$300,000, providing adequate performance to reveal also photospheric line asymmetries and other signatures of stellar photospheric structure.  Also ESPRESSO, for the combined focus of the four VLT unit telescopes of ESO on Paranal \citep{pepeetal14}, combines significant resolution with great light-gathering capability and high wavelength stability.  Observing time on such instruments is virtually guaranteed since the detailed spectroscopy of exoplanets during transit of bright stars is a high-priority project of exoplanet research and the data required for stellar analyses will be obtained concurrently.  In the somewhat more distant future, one can look forward to the planned HIRES instrument at ELT, the ESO Extremely Large Telescope \citep{maiolinoetal13}.  This may not reach higher spectral resolutions than its predecessors, but the order-of-magnitude increase in the collecting area should enable us to reach fainter and rarer stars, perhaps chemically peculiar and magnetic stars. This may also enable the retrieval of spectra from local features, such as starspots, whenever a planet should happen to cross.

\begin{acknowledgements}
{This study used data obtained from the ESO Science Archive Facility, originating from observations made with ESO Telescopes at the La Silla Paranal Observatory under program ID:077.C-0379(A) by I.\ A.\ G.\ Snellen, A.\ Collier Cameron, and K.\ Horne.  At Lund Observatory, contributions to the examination of archival spectra from different observatories were made also by Tiphaine Lagadec and Joel Wallenius.  HGL acknowledges financial support by the Sonderforschungsbereich SFB881 `The Milky Way System' (subproject A4) of the German Research Foundation (DFG).  The work by DD was performed in part at the Aspen Center for Physics, which is supported by National Science Foundation grant PHY-1066293.  DD also acknowledges stimulating stays as a Scientific Visitor at the European Southern Observatory in Santiago de Chile.  We thank the referee for constructive and valuable comments. }

\end{acknowledgements}


\end{document}